\documentclass[aps,pra,twocolumn, notitlepage,10pt,superscriptaddress]{revtex4-1}
\usepackage{graphicx}
\usepackage{times,bbm,amsmath,amssymb}
\usepackage{hyperref}
\usepackage{color}
\usepackage{graphicx}% Include figure files
\usepackage{dcolumn}% Align table columns on decimal point
\usepackage{bm}% bold math
\usepackage{soul}

\usepackage{xcolor}
\usepackage{amssymb}% http://ctan.org/pkg/amssymb
\usepackage{pifont}% http://ctan.org/pkg/pifont
\usepackage{tikz}

\begin{document}

\title{Experimental multiparameter quantum metrology in adaptive regime}

\author{Mauro Valeri}
\thanks{These authors contributed equally}
\affiliation{Dipartimento di Fisica, Sapienza Universit\`{a} di Roma, Piazzale Aldo Moro 5, I-00185 Roma, Italy}

\author{Valeria Cimini}
\thanks{These authors contributed equally}
\affiliation{Dipartimento di Fisica, Sapienza Universit\`{a} di Roma, Piazzale Aldo Moro 5, I-00185 Roma, Italy}

\author{Simone Piacentini}
\affiliation{Istituto di Fotonica e Nanotecnologie, Consiglio Nazionale delle Ricerche (IFN-CNR), Piazza Leonardo da Vinci, 32, I-20133 Milano, Italy}

\author{Francesco Ceccarelli}
\affiliation{Istituto di Fotonica e Nanotecnologie, Consiglio Nazionale delle Ricerche (IFN-CNR), Piazza Leonardo da Vinci, 32, I-20133 Milano, Italy}

\author{Emanuele Polino}
\affiliation{Dipartimento di Fisica, Sapienza Universit\`{a} di Roma, Piazzale Aldo Moro 5, I-00185 Roma, Italy}

\author{Francesco Hoch}
\affiliation{Dipartimento di Fisica, Sapienza Universit\`{a} di Roma, Piazzale Aldo Moro 5, I-00185 Roma, Italy}

\author{Gabriele Bizzarri}
\affiliation{Dipartimento di Fisica, Sapienza Universit\`{a} di Roma, Piazzale Aldo Moro 5, I-00185 Roma, Italy}

\author{Giacomo Corrielli}
\affiliation{Istituto di Fotonica e Nanotecnologie, Consiglio Nazionale delle Ricerche (IFN-CNR), Piazza Leonardo da Vinci, 32, I-20133 Milano, Italy}

\author{Nicol\`o Spagnolo}
\affiliation{Dipartimento di Fisica, Sapienza Universit\`{a} di Roma, Piazzale Aldo Moro 5, I-00185 Roma, Italy}

\author{Roberto Osellame}
\email{roberto.osellame@cnr.it}
\affiliation{Istituto di Fotonica e Nanotecnologie, Consiglio Nazionale delle Ricerche (IFN-CNR), Piazza Leonardo da Vinci, 32, I-20133 Milano, Italy}

\author{Fabio Sciarrino}
\email{fabio.sciarrino@uniroma1.it}
\affiliation{Dipartimento di Fisica, Sapienza Universit\`{a} di Roma, Piazzale Aldo Moro 5, I-00185 Roma, Italy}

\begin{abstract}
Relevant metrological scenarios involve the simultaneous estimation of multiple parameters. The fundamental ingredient to achieve quantum-enhanced performances is based on the use of appropriately tailored quantum probes. However, reaching the ultimate resolution allowed by physical laws requires non trivial estimation strategies both from a theoretical and a practical point of view. A crucial tool for this purpose is the application of adaptive learning techniques. Indeed, adaptive strategies provide a flexible approach to obtain optimal parameter-independent performances, and optimize convergence to the fundamental bounds with limited amount of resources. Here, we combine on the same platform quantum-enhanced multiparameter estimation attaining the corresponding quantum limit and adaptive techniques. We demonstrate the simultaneous estimation of three optical phases in a programmable integrated photonic circuit, in the limited resource regime. The obtained results show the possibility of successfully combining different fundamental methodologies towards transition to quantum sensors applications. 
\end{abstract}

\maketitle

\section*{Introduction}

Quantum correlation has revealed to be a fundamental resource in a large variety of fields ranging from computation and communications to metrology and sensing \cite{giovannetti2004quantum,giovannetti2006quantum,giovannetti2011advances,pezze2009entanglement,pirandola2018advances,polino2020photonic,barbieri2022optical}. In the latter, the use of quantum probes enables enhanced measurement sensitivity with respect to their classical counterpart. Given this paradigm, several classes of quantum sensors such as atomic clocks, magnetic sensors \cite{PhysRevLett.125.020501}, networks of sensors \cite{zhao2021field, proctor2018multiparameter, liu2021distributed,xia2020demonstration,guo2020distributed,ge2018distributed} have been developed. In several practical scenarios, such as imaging and microscopy, the estimation process generally requires the simultaneous measurement of more than one parameter. This consideration motivated a growing interest in investigating multiparameter quantum estimation, from both a theoretical and an experimental perspective \cite{szczykulska2016multi,polino2020photonic, albarelli2020perspective}. 

Several open challenges still need to be addressed to fully exploit the potential of quantum-enhanced estimation in the multiparameter regime. These open points include the design of appropriate strategies to generate the most suitable probes, depending on the specific set of parameters and on the technological peculiarities of the quantum sensor. Then, the quality of the estimation strategies can be assessed studying the Quantum Fisher Information (QFI) \cite{liu2019quantum} from which it can be derived the ultimate precision bound consisting in the quantum Cramér-Rao bound (QCRB) \cite{paris2009quantum}. Such a quantity is valid in the asymptotic resource regime, and depends on the particular probe state chosen to investigate the process and on its interaction with the parameters of interest. Tighter bounds can be evaluated in different regimes, depending also on the available prior information on parameters \cite{demkowicz2020multi,d2022experimental,rubio2019bayesian,cimini2020diagnosing}. After having identified the correct probe, to achieve the ultimate bound, it is necessary to optimize also the adopted measurement strategy.
Furthermore, the realization of an actual quantum sensor needs to face a detailed counting of the number of employed resources. It is then important to optimally allocate them to demonstrate quantum-enhanced sensitivity, independently of the parameter values under investigation.
To this aim, a crucial tool is represented by adaptive strategies which are able to optimize the measurement apparatus parameters during the estimation protocols \cite{wiseman1995adaptive}.   %\textcolor{blue}{On the other hand, the unconditional quantum advantage can be claimed if classical limits are overcome even when all the generated resources, including loss and noise mechanisms, are taken into account \cite{slussarenko2017unconditional,zhao2021field}.} \textcolor{blue}{\textit{[[NS: non sono sicuro sia utile mettere l'accento sul discorso delle losses.]]}}

%\textcolor{blue}{\it{[[NS: Vi proporrei qui di parlare di bound più in generale (non solo di Cramer-bound), in particolare quelli a risorse limitate. Vogliamo secondo me dire che la definizione dei probes viene insieme allo studio dei bounds, in regime di risorse finite (non usiamo limited ma finite), e di informazione a priori arbitraria.]]}}

Multiport interferometry, allowing to investigate multiphases estimation processes \cite{ciampini2016quantum,humphreys2013quantum,polino2019experimental}, is an especially useful platform to develop such methodologies. Some relevant works have been done in this direction \cite{pezze2017optimal,Hong2021} in non-adaptive regimes. However, increasing the number of optical modes, and subsequently the number of phases which can be estimated efficiently, requires to deal with several experimental issues. Indeed, the optimal sensitivity over multiple parameters can be achieved probing the process with high-dimensional entangled states. The realization of such kind of states is still limited to small number of modes. To solve scalability issues, integrated photonics represents an optimal solution \cite{carolan2015universal,wang2020integrated,corrielli2021femtosecond}, allowing to implement complex and tunable transformations on the input states. In particular, integrated circuits permit to easily realize multiport interferometers with the possibility of handling several embedded phases among the different arms. One of the principal strengths of such kind of platforms is the great stability achieved, necessary to implement multiphase estimation protocols. Integrated devices meet almost all the fundamental prerequisites to accomplish quantum-enhanced estimations.
%In particular, integrated photonics permits to realize tunable and scalable devices suitable to accomplish such requirements.
Such devices indeed allow to easily switch the desired input states and the performed measurement schemes, and at the same time they permit a fast tuning of control parameters to implement adaptive protocols.

In this work, we satisfy simultaneously all the aforementioned requirements in a single experiment.
In particular, we report multiparameter estimation of $3$ optical phases, demonstrating experimentally the capability to overcome the optimal separable sensitivity limit, exploiting a two-photon input state with $2$ photons distributed in a 4-arm interferometer. Notably, this is done by employing a Bayesian adaptive protocol that allows to efficiently allocate the number of resources for each estimation, while ensuring an optimized convergence to the ultimate bound in the limited resource regime. Indeed, the application of real-time adaptive feedbacks enable approaching such bound already after only $\sim 50$ probes. This procedure is shown to provide performances which are independent of the particular value of the unknown parameters.
%In this work we combine all the aforementioned open issues in a post-selection configuration demonstrating the experimental saturation of the QCRB for a $3-$parameter estimation problem, exploiting a multimode entangled state with $N=2$ photons and $m=4$ optical modes. Notably, employing a Bayesian adaptive protocol we can efficiently allocate the number of resources for each estimation ensuring a faster convergence to the ultimate bound in the limited resource regime.
Differently from \cite{polino2019experimental,hong2021quantum} we implement an adaptive protocol capable to achieve quantum enhancement in a limited data regime. This kind of protocols has been previously investigated only in classical regime \cite{Valeri2020} and for quantum single-parameter estimation \cite{xiang2011entanglement,berni2015ab,daryanoosh2018experimental}.

Multiparticle input states enable us to perform a multiphase quantum enhanced estimation which, from a conceptual point of view, represents a paradigmatic test bed for multiparameter estimation protocols in the quantum regime. Finally, we compare our results with the ones achievable by probing the system with a sequence of optimal classical probe states, demonstrating an enhancement on the simultaneous estimation of the three phases, surpassing the classical limit and saturating the QCRB.% by 2.3 standard deviations. 
%Moreover, such a task discloses a plethora of applications ranging from.... sensing \cite{}, communication \cite{sidhu2021advances}. 

%\textcolor{blue}{\textit{[[NS: questa parte cosi' non può stare dopo "In this work". Proverei a riusare una buona parte di questo integrandolo nei due paragrafi: (a) ''Multiport interferometry ...'', prendendo la parte che vale come prior art. Di nuovo, non parliamo di entangled states, non e' una cosa che andiamo a declinare nel paper. (b) ''In this work...", mettendoci i punti chiave del nostro paper.]]}}

\subsection*{Multiparameter quantum metrology: multiphase estimation}

%{fisher quantum qcrb - multiphase -quantum multiphase stati ottimali  - problema adattativo tecniche - problema implementazioni sperimentali

A multiparameter approach to quantum metrology has proven to be beneficial in different scenarios where the simultaneous estimation of multiple parameters can provide better precision than estimating them individually by using the same amount of resources \cite{PhysRevLett.116.030801,humphreys2013quantum,szczykulska2016multi,demkowicz2020multi,albarelli2020perspective,ciampini2016scirep}. Note that different strategies and paradigms have been recently considered to quantify the corresponding achievable limits \cite{gorecki2022multiple,gorecki2022multiparameter}. Furthermore, in an actual experiment, even if the parameter of interest is a single one, the estimation process unavoidably involves other parameters, linked to noise, which have to be estimated simultaneously to provide an unbiased estimation \cite{mihai14,Roccia18}.
While in the single parameter scenario the QCRB can in principle be always saturated choosing appropriate measurement schemes, an additional problem arises in the multiparameter case. Here, the saturability of the bound is not always guaranteed \cite{helstrom1976quantum,ragy2016compatibility}. 
It is of particular interest to identify, within such a framework, quantum resources able to obtain a sensitivity advantage versus classical strategies.
The ultimate achievable bound is indeed related to the estimation of the vector $\bm{\varphi}=(\varphi_1,\varphi_2,...,\varphi_d)$ of $d$ parameters becoming an inequality on their covariance matrix:
\begin{equation}
    \Sigma(\bm{\varphi})\ge \frac{\mathcal{F _C}^{-1}(\bm{\varphi})}{M}\ge\frac{\mathcal{F _Q}^{-1}(\bm{\varphi})}{M}\; ,
\label{eq:CRB}    
\end{equation}
where  the covariance matrix is given by:
\begin{equation}
    \Sigma(\bm{\varphi})_{ij}=\sum_{\bm{x}}[\bm{\hat{\varphi}}(\bm{x})-\bm{\varphi}]_{i}\:[\bm{\hat{\varphi}}(\bm{x})-\bm{\varphi}]_{j} \; P(\bm{x}|\bm{\varphi}).
\end{equation}
Here $i,j=1,...,d$, $\bm{\hat{\varphi}}$ is the list of estimators of $\bm{\varphi}$, $\bm{x}$ are the possible outcomes,  $P(\bm{x}|\bm{\varphi})$ is the likelihood of the estimation process. In the inequalities $\mathcal{F _C}$ is the Fisher Information matrix defined as: $\mathcal{F _C}(\bm{\varphi})_{ij}=\sum_x\left[\frac{1}{ P(x|\bm{\varphi})}\frac{\partial P(x|\bm{\varphi})}{\partial\varphi_i}\frac{\partial P(x|\bm{\varphi})}{\partial\varphi_j}\right]$, $M$ is the number of measurements while  $\mathcal{F_Q}(\bm{\varphi})$ is the quantum Fisher Information (QFI) matrix. The first inequality is referred to as the Cramér-Rao bound (CRB) while the second inequality, i.e. QCRB, in the multiparameter scenario is fulfilled only if the collective saturation of the bound for all the parameters is simultaneously verified \cite{PhysRevLett.127.110501}. Therefore, of particular interest are those situations where the optimal measurement schemes for each parameter are compatible and consequently the right hand of the inequality \eqref{eq:CRB} becomes an equality, making the CRB equal to the QCRB. Such bounds are relative to the frequentist approach \cite{li2018frequentist}  where the parameter is approximated with the estimator that usually coincides with the one maximizing the likelihood of the measurement results. The sensitivity of the multiparameter estimation can be obtained by computing the trace of the covariance matrix, which is then compared with the trace of the FI and of the QFI. Note that this is not the only possible choice for the definition of sensitivity. Indeed, different figures of merit can be used, such as sums of FI terms with general weights \cite{malitesta2021distributed}.

The saturation of the QCRB is verified asymptotically, therefore, in a real scenario where only a limited amount of resources is available, it is important to optimize them at each step in order to ensure the convergence. The optimization of the resources can be implemented through adaptive strategies which indeed ensure a faster convergence to the ultimate bound. Adaptive Bayesian estimation protocols are usually employed to accomplish such tasks \cite{KitLee,Valeri2020} where at each step the posterior distribution is updated depending on the settings of some control parameters. Although the aforementioned bounds are not computed for Bayesian estimation, in the limit of a large number of repeated measurements the frequentist and the Bayesian methods agree, therefore the QCRB can still be  employed as a reference for Bayesian settings in the asymptotic regime.

One of the most investigated framework for studying multiparameter estimation is optical interferometry, where the unknown parameters are mapped in the different phase shifts between the arms of an interferometer with respect to a reference \cite{macchiavello2003optimal,humphreys2013quantum,liu2016quantum,gagatsos2016gaussian,ciampini2016quantum,pezze2017optimal,gessner2018sensitivity,oh2020optimal,polino2019experimental,Valeri2020,Hong2021,goldberg2020multiphase,gebhart2021bayesian,gorecki2022multiple,d2022experimental}. The quantum estimation of phases is of paramount importance for different applications: apart from direct use in sensing like biological imaging \cite{taylor2013biological}, it can be employed also in tasks such as quantum communication \cite{rudolph2003quantum}, %quantum sampling \cite{temme2011quantum},
simulation \cite{o2016scalable} and even gravitational wave detection \cite{abadie2011gravitational}.

Lastly, and importantly, multiphase estimation is a paradigmatic scenario representing a fundamental test bed for general multiparameter estimation protocols. In this context, a probe  $|\Psi_0 \rangle$, prepared by a suitable operation in the space of $d+1$ modes, interacts with the phase shifts through the unitary evolution: 
$|\Psi_{\bm{\varphi}}\rangle=e^{(i\sum_{i=1}^{d}n_i\varphi_i)} |\Psi_0 \rangle$, 
 where $n_i$ is the generator of the phase $\varphi_i$ along the mode $i$, i.e. the photon number operator for that  mode. Since such generators commute, $[n_i, n_j]=0$ $\forall i,j$ the QFI matrix 
$\mathcal{F _Q}(\bm{\varphi})_{ij}=4 [\langle n_i n_j\rangle-\langle n_i\rangle \langle n_j\rangle]$, where the average $\langle\cdot\rangle$ is over $|\Psi_{\bm{\varphi}}\rangle$ and the probe states are assumed to be pure \cite{polino2020photonic}.
Finally, after a final transformation, the state is measured and an estimator provides the estimation of the unknown phases.
It has been demonstrated that the optimal quantum probe state, together with the optimal measurement, can achieve quantum enhanced performances with also an advantage of order $O(d)$ over the best quantum precision for the phases estimated separately \cite{humphreys2013quantum}. Note that the improvement $O(d)$ achieved by the simultaneous estimation is reduced to a constant if the resource count is chosen differently \cite{gorecki2022multiple}.

Particular interest needs to be devoted to identify those configurations, i.e. number of optical modes constituting the arms of the multiport interferometer and the possible input states, which allow to saturate the ultimate bound of precision \cite{ciampini2016scirep}. These configurations demonstrate enhanced performances compared to the use of classical probe states. In particular, states having a coherent superposition of $M$ photons in one modes and none in the others, allowing the simultaneous estimation of multiple phases, achieve advantaged performances compared to any classical probe states. The need of controlling the input states, as well as the performed measurements and configuring some control parameters to implement adaptive protocols requires a versatile and programmable platform. All these conditions are attained by integrated photonics which represents a promising platform for quantum sensing and metrology studies and applications \cite{wang2020integrated}.

%overcomes optimal separable strategies for the estimation of any linear combination of the phases
%The sensing scheme can follow an entangled (also indicated as parallel, or global, in the literature) or a separable (sequential, or local) strategy

\section*{Results}
\subsection*{Integrated multiport interferometer for quantum sensing}
%5 righe su circuiti integrati potenzilità e applicazioni contour plot 3D con stellina

\begin{figure*}[ht!]
\centering
\includegraphics[width=0.99\textwidth]{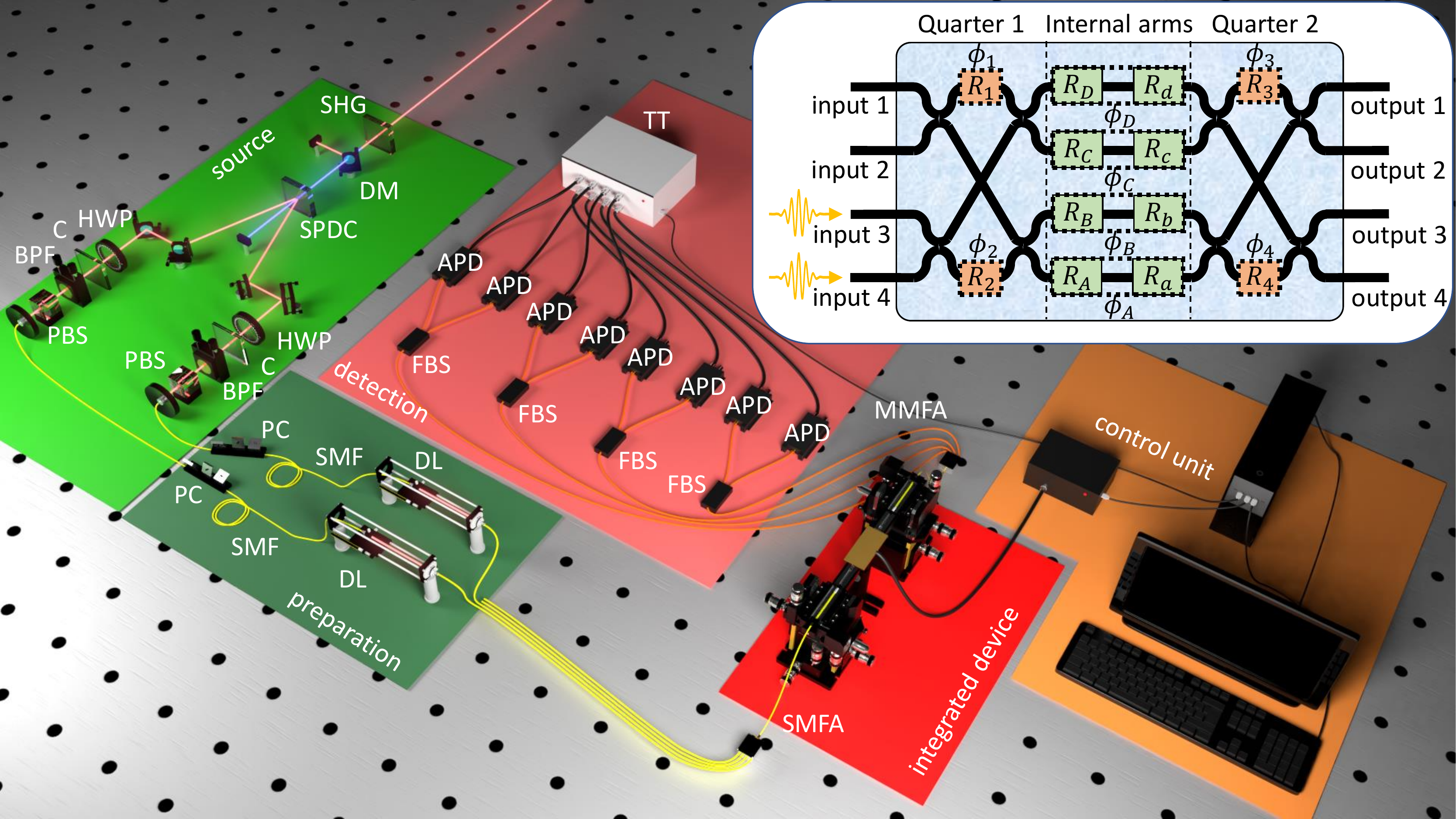}
\caption{\textbf{Full experimental setup employed for multiphase estimation experiments on chip.} (i) Source. Two-photon states are generated via parametric-down-conversion in a BBO ($\beta$-Barium Borate) crystal. (ii) Preparation. Photons are made indistinguishable in time via delay lines, and in the polarization degree of freedom via fiber polarization controller. (iii) Integrated device. Photons are injected in the integrated interferometer, and collected at the output, via fiber-arrays. (iv) Detection. Two-photon events are collected via a probabilistic photon-counting scheme. (v) Control unit. Measurement outcomes are processed by the control unit, and are used to drive the thermal shifter operation. Inset: Scheme of the integrated circuit. Each thermal shifter $R_i$ is able to control a specific optical phase of the device. In particular, with regard to the internal arms of the interferometer, we identify the three independent internal phase-shifts $(\varphi_A,\varphi_B,\varphi_D):=(\phi_A-\phi_C,\phi_B-\phi_C,\phi_D-\phi_C)$ by setting $C$ as reference arm. Instead, $\phi_{1-4}$ define the equivalence class of each quarter transformation. Legend. SHG: second harmonic generation. DM: dichroic mirror. SPDC: spontaneous parametric down conversion. HWP: half-wave plate. C: walk-off compensation. BPF: band-pass filter. PBS: polarizing beam-splitter. PC: polarization compensation. SMF: single-mode fiber. DL: delay line. SMFA: single-mode fiber array. MMFA: multi-mode fiber array. FBS: fiber beam-splitter. APD: avalanche photodiode. TT: time-tagger.}
\label{fig:setup}
\end{figure*}

Our platform consists in an actively tunable integrated 4-arm interferometer realized through femtosecond laser waveguide writing in glass \cite{corrielli2021femtosecond, meany2015laser}. In particular, the device is composed by two cascaded quarters, which are $4\times4$ optical elements that split equally the optical power at all its input ports across all output ports. Each quarter is composed by four directional couplers arranged in a two-layers configuration and a three-dimensional waveguide crossing, as depicted in Fig.~\ref{fig:setup}. Moreover, each quarter is equipped with two thermal phase shifters ($R_{1-4}$), which allow to actively control the internal optical phase between the directional coupler layers ($\phi_{1-4}$), and select a specific equivalence class of the quarter transformations \cite{zukowski1997realizable}. Between the two quarters, the interferometric region is composed by four straight waveguide segments whose optical phases $\phi_{A-D}$ can be controlled by means of $8$ thermal phase shifters ($R_{a-d}$ and $R_{A-D}$). The overall length of the device is $3.6$ cm. All thermal shifters have been fabricated by femtosecond laser micromachining and include laser-ablated isolation trenches around each microheater \cite{ceccarelli2020low}. This configuration allows to both reduce the power consumption ($2\pi$ phase shift on a single resistor is obtained by dissipating less than $25$ mW of electrical power) and to greatly reduce the thermal cross talk between adjacent shifters. More details regarding the circuit geometry, the waveguide inscription, the thermal shifter fabrication processes and the thermal shifter performances are provided in Supplementary Note 1. Finally, two $4-$channels single mode fiber arrays have been glued at the interferometer input and output facets, with average fiber-to-fiber total insertion losses (from the connector of the input fiber to the connectors of the output fiber array) of 2.5 dB
(insertion loss of the bare device before pigtailing of 1.5 dB).

On the basis of the presented scheme, the transformation performed by the phase shifters fabricated on the internal arms of the interferometer reads:
\begin{equation}
U_\phi = 
    \begin{bmatrix}
e^{i\phi_D} & 0 & 0 & 0\\
0 & e^{i\phi_C} & 0 & 0\\
0 & 0 & e^{i\phi_B} & 0\\
0 & 0 & 0 & e^{i\phi_A}\\
\end{bmatrix}
,
\end{equation}
while the relation linking the dissipated power $\omega$ to the inserted phase shift can then be approximated by:
\begin{equation}
    \varphi_i = \sum_j (\alpha_{ij}\omega_j + \alpha^{(2)}_{ij}\omega_i\omega_j) + \varphi_{0i},
\label{phases}
\end{equation}
where $\varphi_0$ is the zero-current phase shift, while $\alpha$ and $\alpha^{(2)}$ are respectively the linear and quadratic response coefficients associated to the phase shift $\varphi$.
In particular, in our device $12$ thermo-optic phase shifters can be suitably controlled. The interferometer is able to perform the simultaneous estimation of three independent phase shifts $\bm{\varphi}$ between three arms and a reference one. In the following, we choose $C$ as reference arm, thus considering $(\varphi_A,\varphi_B,\varphi_D)\equiv(\phi_A-\phi_C,\phi_B-\phi_C,\phi_D-\phi_C)$ as the triple phases to be estimated. 
The transformation induced by the actual device will also depend on the effective trasmittivities and reflectivities of the $8$ directional couplers. 

We start by theoretically studying the operation and the bounds relative to the ideal device i.e. when the reflectivities and trasmittivities of all the directional couplers are equal to the nominal value of $\frac{1}{2}$. The QFI depends only on the prepared probe state, therefore it is a function of the input modes of the injected photons and of the phases $\phi_1$ and $\phi_2$ of the first quarter whose transformation is given by: 
\begin{equation}
U_Q = \frac{1}{2}
    \begin{bmatrix}
e^{i\phi_2} & ie^{i\phi_2} & i & -1\\
ie^{i\phi_2} & -e^{i\phi_2} & 1 & i\\
i & 1 & -e^{i\phi_1} & ie^{i\phi_1}\\
-1 & i & ie^{i\phi_1} &e^{i\phi_1}\\
\end{bmatrix}
.
\label{quarter}
\end{equation}
However, depending on the specific input the dependence on these two phases can vanish. More specifically this condition is verified when injecting two photons either in the first two modes ($|1100\rangle$) or in the last two ($|0011\rangle$). Such a choice allows to generate, after the first quarter, the  multiphoton entangled input state:

\begin{equation}
\begin{split}
|\psi_0\rangle &= \frac{i}{2\sqrt{2}}(|2000\rangle - |0200\rangle + e^{-2i\phi_1} |0020\rangle +\\
&-e^{-2i\phi_1} |0002\rangle) - \frac{1}{2}(|1100\rangle + e^{-2i\phi_1} |0011\rangle).
\end{split}
\label{stato_in}
\end{equation}
For our device, the use of two-photon quantum probes ensures to approach the ultimate asymptotic quantum limit for the $3-$phase estimation represented by the relative QCRB which results to be equal to $2.5/M$. The computed bound represents the ultimate quantum limit achievable in the estimation precision for the considered input.

The optimality of the full scheme is therefore demonstrated when the CRB, obtained after the measurement process is also considered, reaches the QCRB. Therefore, when studying the CRB also the characteristics of the second quarter must be considered in the model. The state generated at the output after injecting into the device two photons in the third and fourth input is a coherent superposition of $2$ photons in the $4$ output modes of the device: 
\begin{equation}
\begin{split}
|\psi\rangle_{out} &= a_{11}|2000\rangle + a_{22}|0200\rangle + a_{33}|0020\rangle + a_{44}|0002\rangle +\\ &+ a_{12}|1100\rangle + a_{13}|1010\rangle + a_{14}|1001\rangle + a_{23}|0110\rangle +\\ &+ a_{24}|0101\rangle + a_{34}|0011\rangle
\end{split}
\label{stato}
\end{equation}
with $a_{11} = a_{22}$, $a_{33} = a_{44}$, $a_{13} = a_{24}$ and $a_{23} = a_{14}$ where all the coefficients now depend on the parameters imposed by $U_Q$ transformation and on the particular settings of $\phi_1$, $\phi_2$, $\phi_3$ and $\phi_4$.
The CRB, given such a state, can indeed saturate the ultimate limit of $2.5/M$, satisfying the general necessary conditions for the saturation of QCRB of multiphase estimation in interferometric setups \cite{pezze2017optimal}. 
It is fundamental to notice that indistiguishability between the two input photons is a necessary condition to reach such bound. The minimum of the CRB in the scenario of indistinguishable photons ensures the saturation of the QCRB, and the achievement of a quantum enhanced estimation over $3$ parameters. Indeed, the use of completely distinguishable photons allows to achieve a minimum equal to $3$ (see Supplementary Note 2 for details). 

To demonstrate the capability of reaching an estimation enhancement, we compare our result also with the optimal estimation obtained through single-photon states \cite{humphreys2013quantum}. 
%It is known that the most efficient multiphase estimation can be achieved with a probe consisting in a coherent superposition of $N$ photons in one of the modes and none in any of the others allowing a simultaneous estimation of the parameters of interest \cite{humphreys2013quantum}. 
The achievable bound using two of such optimal single-photon states on our system is $\text{QCRB}=2.8$.
Therefore, the saturation of $2.5/M$ demonstrates quantum-enhanced measurement sensitivity reachable with indistinguishable two-photon states compared to any sequence of classical single-photon probes and independent measurement, even including the optimal single-photon state.

%\begin{figure}[ht!]
%\centering
%\includegraphics[width=.5\textwidth]{QCRB_ideal_1.PNG}
%\caption{QCRB for input states: $|1010\rangle$,$|0101\rangle$,$|0110\rangle$ and $|1001\rangle$.}
%\label{fig:QCRB} 
%\end{figure}

\begin{figure}[ht!]
\centering
\includegraphics[width=0.5\textwidth]{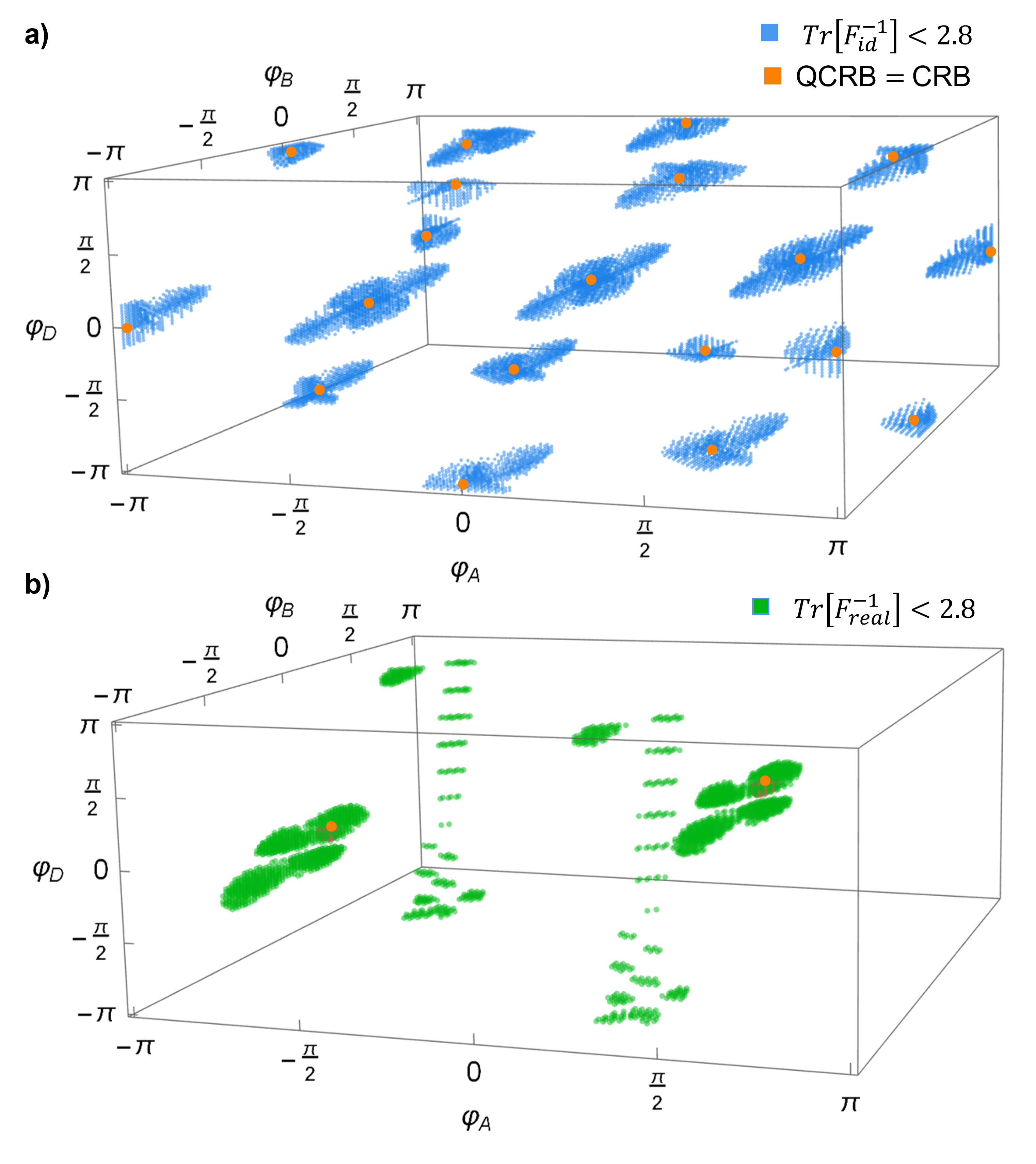}
\caption{\textbf{Cramér-Rao Bound regions.} Points corresponding to a value of the CRB $<2.8$. The orange points correspond to the minimum where the QCRB is saturated. \textbf{a)} Bound relative to the ideal device. \textbf{b)} Bound for the real device whose minimum is $2.53/M$.} %one is the bound computed with all the parameters of the real device while the blue region is computed with the ideal device. The green star represents the minimum of the CRB for the real device.
\label{fig:volume} 
\end{figure}

%For such device, the use of two-photon quantum probes ensures to approach the ultimate quantum limit for the $3-$phase estimation. The saturation is guaranteed when the Cramér-Rao bound (CRB) after the measurement process reaches the QCRB. To study the CRB also the characteristics of the second quarter must be considered in the model. It is fundamental to notice that the necessary condition to achieve the ultimate limit is the indistiguishability of the two input photons. Such condition can be easily verified looking at Fig.\ref{fig:CRB} where the CRB obtained when employing $2-$photon indistinguishable and completely distinguishable states are reported. The minimum of the CRB in the scenario of indistinguishable photons ensure the saturation of the QCRB and the achievement of a quantum enhanced estimation over $3$ parameters. Indeed, the use of complete distinguishable photons allows to achieve a minimum equal to $3$. The optimal estimation, on our system, using non entangled states is QCRB$=2.8$ given by ... \textcolor{blue}{(discorso su stati di Peter)} .
The parameter (phases) regions showing such an advantage where the achieved CRB, for the ideal device, is lower than $2.8$ are limited and are reported in Fig.~\ref{fig:volume}a). However, thanks to the implementation of an adaptive protocol we are able to demonstrate the sensitivity enhancement independently of the values of the estimated triplet of optical phase shifts in the limited resources regime.

%In a second step, after performing a punctual characterization of all the parameters describing the operation of the real device, which allows to reconstruct the likelihood of the system, we can compare the achieved estimation performances with the actual bounds. From this follows that the QCRB will be slightly different

\subsection*{Experimental saturation of the ultimate quantum Cramér-Rao bound}

%Fisher crb e qcrb real saturazione  plot fisher 2 D con star quantum e 3D volumetto 

In order to investigate the actual capabilities of the employed device with two-photon input states, it is necessary to reconstruct its likelihood function through a calibration procedure (see Supplementary Note 3 for details). This step is necessary to derive the achievable CRB with the actual device. 

We reconstruct the $10$ two-photon output probabilities by fitting the measured data for different values of voltages applied to the resistors of the device. In particular, we collect measurements studying the device response as a function of the power dissipated on the three thermal shifters, i.e. $R_a,R_b,R_d$, allowing the complete tuning of the internal phases. In this way, using equations \eqref{phases}, we can model also the effect that the voltage applied on a certain resistor has on the other arms of the device, retrieving all the different cross-talks among the resistors. More specifically, we measure the coincidence events registered at the output of the integrated circuit by dissipating through each selected resistor $10$ different power values, which are equally spaced over the allowed range. More technical details on characterization data can be found in Supplementary Note 3. 

\begin{figure*}[ht!]
\centering
\includegraphics[width=\textwidth]{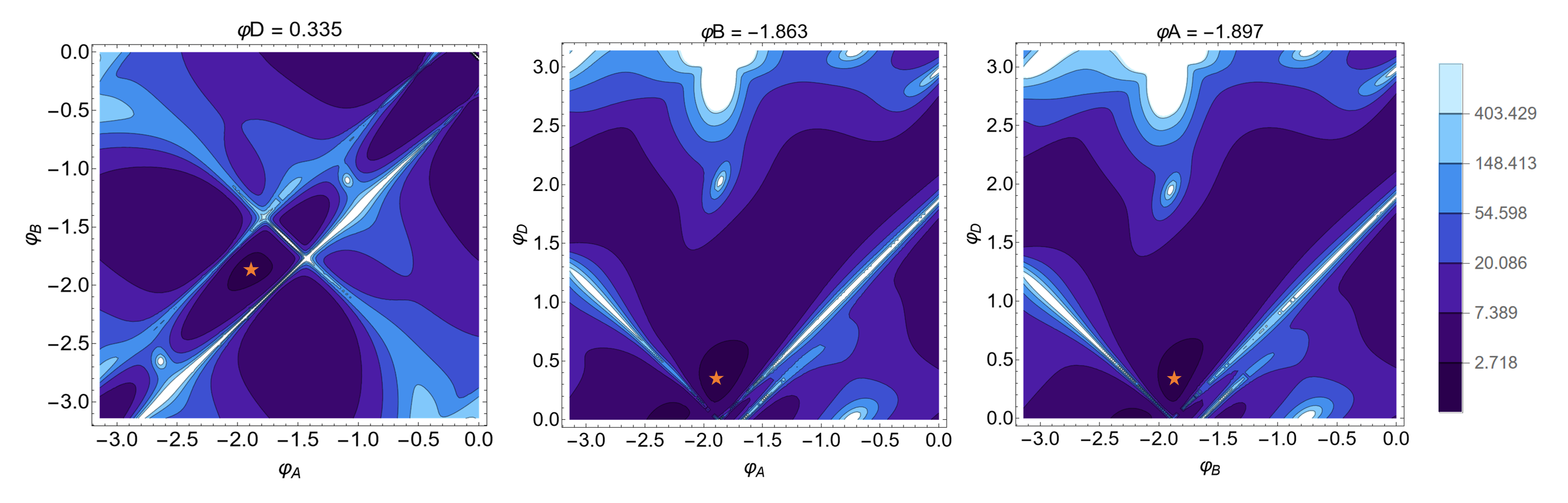}
\caption{\textbf{Slices of Fisher Information matrix.} Three cuts of $\mathrm{Tr}[F^{-1}]$ obtained fixing, from left to right, respectively the values of $\varphi_D$, $\varphi_B$ and $\varphi_A$. The plot shows the sensitivity of different regions of the parameters space.  The orange star represents the minimum of the variance where the CRB is equal to $2.53/M$. }
\label{fig:CRB_real} 
\end{figure*}

Finally, the output probabilities reconstructed from experimental data can be used to compute the FI matrix and to retrieve the experimental CRB. In Fig.~\ref{fig:volume}b we report the regions showing a lower bound compared to the minimum achievable with the best classical states for such calibration. In addition, the inverse of the trace of the FI is reported in Fig.~\ref{fig:CRB_real}. For the actual device, considering all the experimental imperfections, the minimum which corresponds to the achievable bound is $2.53/M$ and it is achieved in two different points of the space (see Fig.~\ref{fig:volume}b). This value is very close to the ideal one of $2.5/M$ and it is still below the critical threshold of $2.8$. With our device we demonstrate quantum enhancement in the simultaneous estimation of three optical phases, experimentally approaching the QCRB in a post-selected configuration.
%riferimento figura 2b --> adattativo

\subsection*{Comparison with the sequential bound}
\label{sec:comparison}

In a general scenario, we can study the sensitivity performances obtained when estimating a linear combination of the parameters under study. Distributed sensing \cite{zhao2021field, proctor2018multiparameter, liu2021distributed,xia2020demonstration,guo2020distributed,ge2018distributed} represents indeed a field that is increasingly being investigated lately. However, instead of looking at any generic combination of parameters $\bm{\nu}\cdot\bm{\varphi} = \sum_{i=1}^d \nu_i\varphi_i$, here, following \cite{malitesta2021distributed}, we can study the achieved performances over the optimal combination of phases to show quantum-enhanced sensitivity. Therefore, we compare for our setup the sensitivities reached with the simultaneous multiparameter estimation with respect to sequential strategies where the different parameters are all estimated independently. 
In particular, the optimal vector $\bm{\nu}$ for our setup is the eigenvector of the QFI matrix associated to the largest eigenvalue $f_{max}$ i.e $\bm{\nu}_{max} = (1/2,1/2,-1/\sqrt{2})$. It follows that the optimal linear combination of optical phases that we can estimate is: $(\phi_A-\phi_D)/2+(\phi_B-\phi_D)/2-(\phi_C-\phi_D)/\sqrt{2}.$ The study of this figure of merit allows to consider also the off-diagonal terms of the QFI that in general depend on mode entanglement in the probe state. 
It is then possible to compute the sensitivity bound on the estimate of the linear combination, achieved when using the employed entangled input states, which is given in \cite{malitesta2021distributed} and results to be:
\begin{equation}
    \Delta^2(\bm{\nu}_{max}\cdot\bm{\varphi}) = \bm{\nu}_{max}\mathcal{F_Q}\bm{\nu}_{max}^T = 0.292.
\end{equation}
The comparison can be done with the optimal separable strategy achieved using coherent states with an average number of photons $\langle\bar{n}\rangle=2$ to estimate sequentially three optical phases embedded in a network of Mach-Zehnder interferometers. In such setting, the QCRB is:
\begin{equation}
    \Delta^2(\bm{\nu}_{max}\cdot\bm{\varphi})_{\text{seq}} = \sum_{i=1}^3 \frac{\nu_i^2}{\mathcal{F}_i}.
\end{equation}
Here, $\mathcal{F}_i$ is the single-parameter QFI for coherent states injected into a Mach-Zehnder interferometer, i.e. $\mathcal{F}_i=\bar{n}_i$ \cite{pezze2008mach}. By numerical optimization, we obtain the minimum of $\Delta^2(\bm{\nu}_{max}\cdot\bm{\varphi})_{\text{seq}} =1.45$, corresponding to the bound achievable with sequential classical measurements. Consequently, a sensitivity on the estimation of the optimal linear combination below this separable bound is a demonstration of the enhancement achieved using entangled probes \cite{malitesta2021distributed}.

\subsection*{Adaptive three-phase estimation}

\begin{figure}[h!]
\centering
\includegraphics[width=0.5\textwidth]{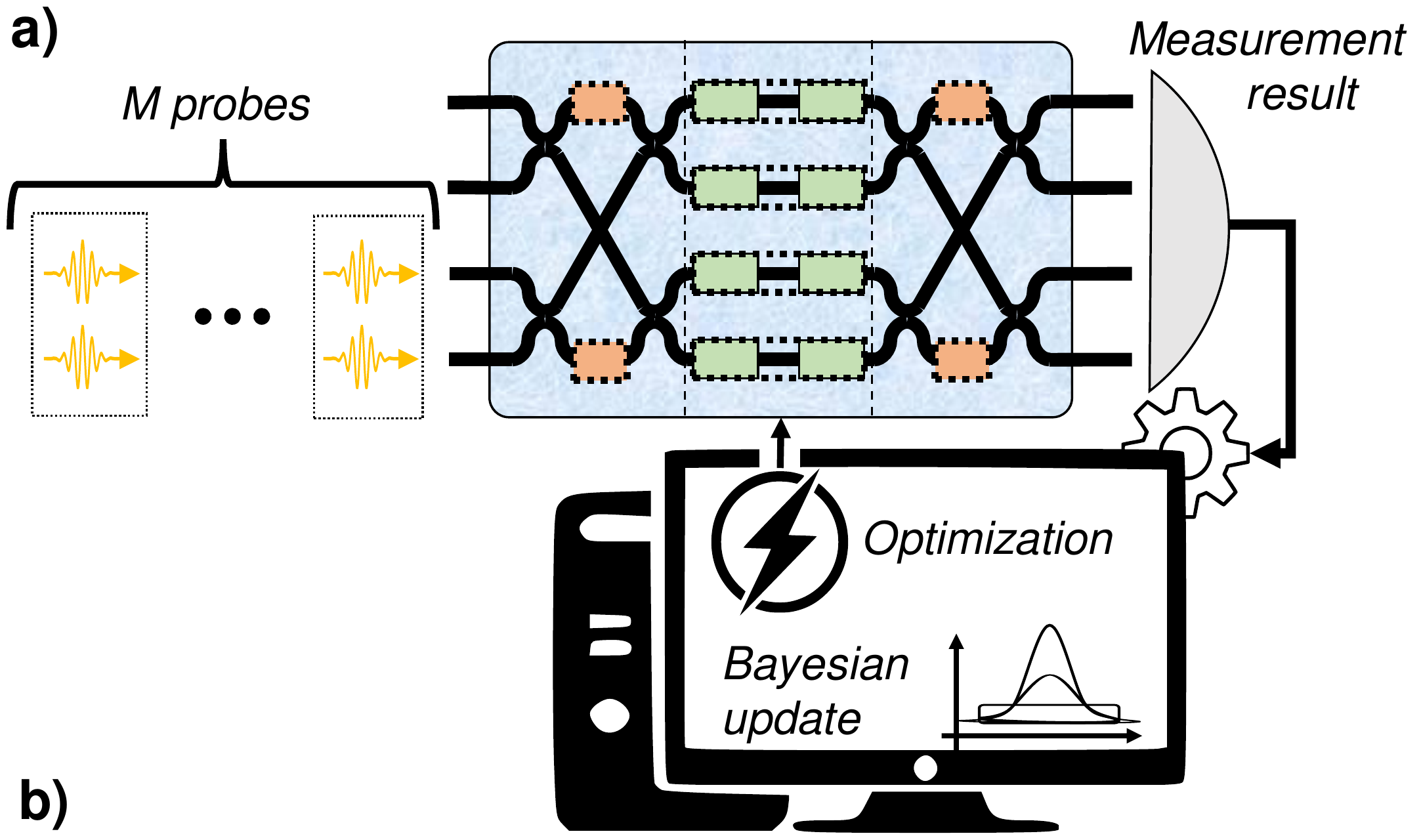}
\includegraphics[width=0.5\textwidth]{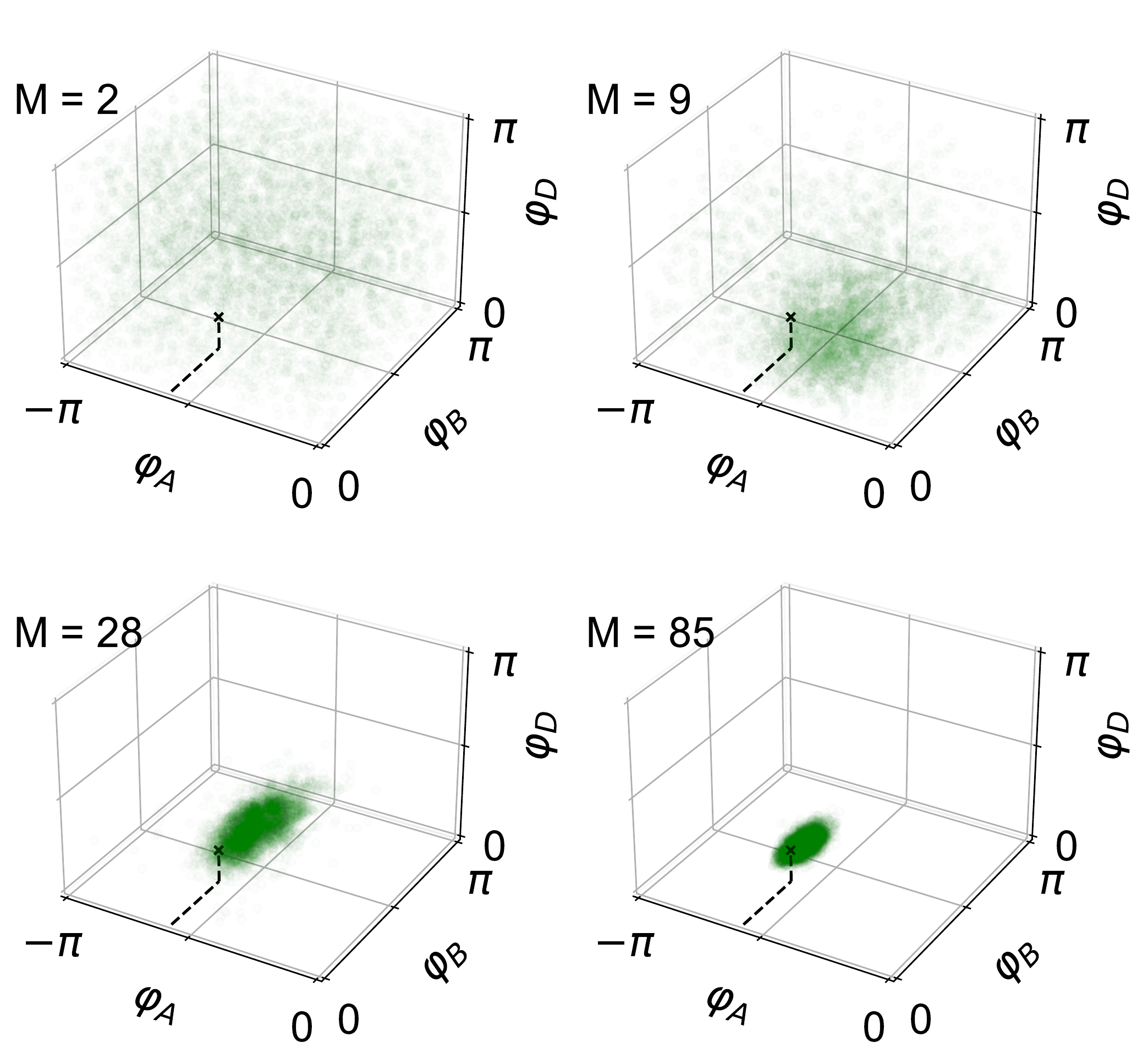}
\caption{\textbf{Adaptive estimation and control feedbacks}. Example of adaptive multiparameter Bayesian learning by injecting a series of $M$ probes into the device. $\textbf{a)}$ After each measurement result, the algorithm computes the best control parameters, i.e. a set of currents to apply for optimizing the estimation efficiency of the next probe. At the same time, each measured probe updates the knowledge of the parameters according to Bayes' rule, concentrating the probability distribution around the true values. $\textbf{b)}$ Evolution of posterior knowledge (green cloud) after sending 2, 9, 28 and 85 probes to estimate the 3 phases ($\varphi_A^{(X)},\varphi_B^{(X)},\varphi_D^{(X)})=(-1.82,1,0.54)$ simultaneously. The distribution converges rapidly around the true values of the triple phases (black cross).}
\label{fig:posterior} 
\end{figure}

Finally, we study the performances achieved when implementing adaptive strategies, able to set the device in the optimal working point for the estimation \cite{wiseman1995adaptive,KitLee,Valeri2020}. This optimization can be done before each probe and it is independent of the specific unknown values. It is based on controlling additional parameters, used as feedbacks during the estimation cycle (Fig.~\ref{fig:posterior}a). Adaptive techniques are used when the number of resources is limited or to solve estimation ambiguities related to the output probability of the system. The capability of asymptotic saturation of lower bounds is not sufficient when an optimal estimation in a few probes is required. Moreover, the computation of which optimal feedbacks have to be applied is in general non-trivial, especially for increasing complexity of the system. For this reason, machine learning techniques are often adopted, able to tackle this hard computational task and in general to enhance sensing protocols \cite{polino2020photonic,vernuccio2022artificial,nolan2021machine,fiderer2021neural,cimini2019calibration,cimini2021calibration,xiao2022parameter}.

Here, we employ a Bayesian framework (see the Supplementary for details) for the adaptive protocol, which represents a powerful tool for multiphase estimation \cite{gebhart2021bayesian,granade2012online}.
In particular, we use the Bayesian multiparameter estimation protocol employed in \cite{wiebe2016efficient,granade2012online,Valeri2020}. Simultaneous adaptive two-phase estimation experiments have been demonstrated without quantum enhancement, injecting a three-mode interferometer with single-photon states \cite{Valeri2020}. Thus, we select such an approach for our multiphase estimation problem demonstrating the saturation of the ultimate precision bounds. 

The realization of adaptive multiphase estimation requires the identification of unknown and control parameters. The structure of our platform allows us to handle independently two layer of internal phases by simply acting on different resistors: the phases to be estimated $\bm{\varphi}^{(X)}$ and the phases to be tuned for adaptive estimation $\bm{\varphi}^{(C)}$, such that $\bm{\varphi}=\bm{\varphi}^{(X)}+\bm{\varphi}^{(C)}$. In our case, the triplet of unknown parameters $\bm{\varphi}^{(X)}$ are set using the thermal shifters $R_A,R_B,R_D$, while the control parameters $\bm{\varphi}^{(C)}$ are tuned using $R_a,R_b,R_d$. To easily achieve adequate control for each estimate, the calibration of resistors $R_a,R_b,R_d$ can be repeated for each selected triplet $R_A,R_B,R_D$. This method guarantees also a more precise calibration of the specific working point of the device.
%For the experiment we used the same setup of the calibration (Fig~?).

The algorithm is based on a sequential Monte Carlo (SMC) technique and it is discussed in detail in the Supplementary material. The SMC guarantees high performances in computing integrals -- replaced by sums -- also when the dimensions of the space increase. The quality of the approximation can be improved by adding further particles, at the cost of a more expensive computation. Then the algorithm allows the computation of the control parameters to be applied during the adaptive estimation. Such optimal values are those which maximize the expected overall variance after measurement of the subsequent probe. Here, the expectation value is computed using the SMC approach.
In order to identify appropriate values of the algorithm parameters for the experiment, we simulated adaptive multiphase estimations for different configurations of such parameters. 
%The simulation results and all the technical details are provided in Supplementary Information. 
A set of phase triplets $\{ \bm{\varphi}^{(X)} \}$ is uniformly selected in $[0,2\pi]\times[0,2\pi]\times[0,2\pi]$ and estimated by a series of two-photon states. The estimation of each triplet is repeated $M=30$ times. A single experiment of adaptive three-phase estimation is reported in Fig.~\ref{fig:posterior}b, by showing how the updated posterior distribution converges to the true value after sending $2, 9, 28$ and $85$ probes. %The output probability distribution of our device presents periodicity of $\pi$ for many internal phases. 
The output probability distribution of our device, given the considered entangled input state, can estimate unambiguously each of the three phases in a $\pi$ range. For this reason, we set the a-priori Bayesian distribution equal to a uniform distribution with a $\pi$ width. 
Note that, by repeating the estimation procedure several times, we obtain the mean of the Bayesian posterior distribution, from which we retrieve the achieved sensitivity for all the performed repetitions, allowing us to compare our results with the bounds of the frequentist scenario.

%thus allowing the algorithm to solve faster this  ambiguity for all values of the unknown phases. Unlike non-adaptive strategies, we note that this is not a necessary step for adaptive approach, which is able to distinguish such ambiguity issues.

%The output probability distribution of our device presents periodicity of $\pi$ shifts for many internal phases, as well as its partial derivative. For the sake of simplicity, we set the a-priori distribution to $\pi$, thus allowing the algorithm to solve this ambiguity for all values of the unknown phases. We note that this is not a necessary step for those values that have the same output probability but different derivatives. Indeed unlike non-adaptive strategies, in this case the adaptive algorithm is able to distinguish the ambiguity issues.

\begin{figure}[htb!]
\centering
\includegraphics[width=0.5\textwidth]{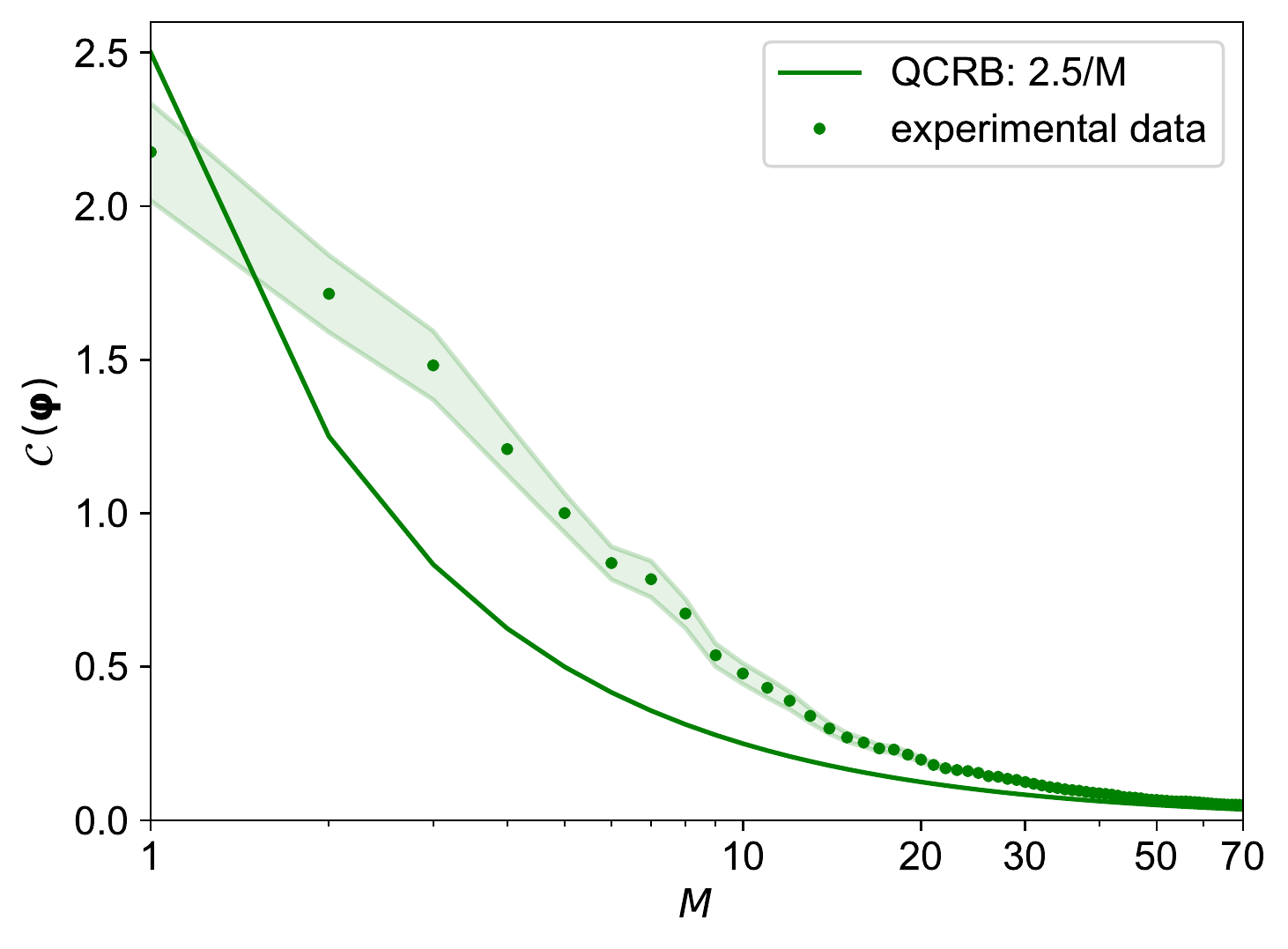}
\caption{\textbf{Experimental adaptive 3-phase estimation.} Quadratic loss $\mathcal{C}(\bm{\varphi})$ is plotted as a function of the number $M$ of injected two-photon input states $|0011\rangle$. Green dots show the performance averaged on 12 different triple phases, estimated using the online Bayesian adaptive technique described in the main text. The experiment for each phase triplet is repeated $30$ times and the final performance is characterized by the mean estimator, the shaded green area is the one standard deviation region. The dashed line is the ultimate precision bound i.e. the QCRB ($2.5/M$) for the ideal device when injected with indistinguishable photons. }
\label{fig:exp} 
\end{figure}

The accuracy of the estimation can be computed looking at different figures of merit. We start investigating a commonly employed one in the first studies of multiphase estimation \cite{humphreys2013quantum} by firstly considering a figure of merit that takes into account the trace of the covariance matrix. Then, we generalize the discussion considering also the off-diagonal terms of the covariance matrix, when demonstrating quantum-enhanced sensitivity for the estimate of a linear combination of the considered parameters. The covariance of the posterior distribution $\Sigma(\bm{\hat{\varphi}})$ represents the confidence interval of the estimate and thus the actual error of the quantum sensor employed. In parallel, the quadratic loss distance $\mathcal{C}(\bm{\hat{\varphi}})$, between the estimated parameters and their true values, provides a reliable evaluation of both the estimation uncertainty and the presence of possible biases. 
In the asymptotic regime the average of both quantities must saturate the CRB. Here, we employ the adaptive technique in order to approach the ultimate precision bound with the minimum number of probes, reporting the experimentally attained quadratic loss function averaged over 12 different triplet of phases. As shown in Fig.~\ref{fig:exp} we are able to reach performances close to the asymptotic limit already after sending around $50$ probes.  %Fig.? reports the QCRB saturation for the median on the $30$ repetitions of one of the measured triple phases. The choice of the median allows to give less weight to the edge events present both for experimental and programming limits, providing a more truthful result. Furthermore, asymptotic convergence is demonstrated for both the mean and the median since asymptotically the two estimators must coincide. 

%\begin{figure*}[hb!]
%\centering
%\includegraphics[width=\textwidth]{Fisher.png}
%\caption{Purple: $F_{11}$, Blue: $F_{22}$, Orange: $F_{33}$}
%\label{fig:fisher} 
%\end{figure*}

\begin{figure}[h!]
\centering
\includegraphics[width=0.5\textwidth]{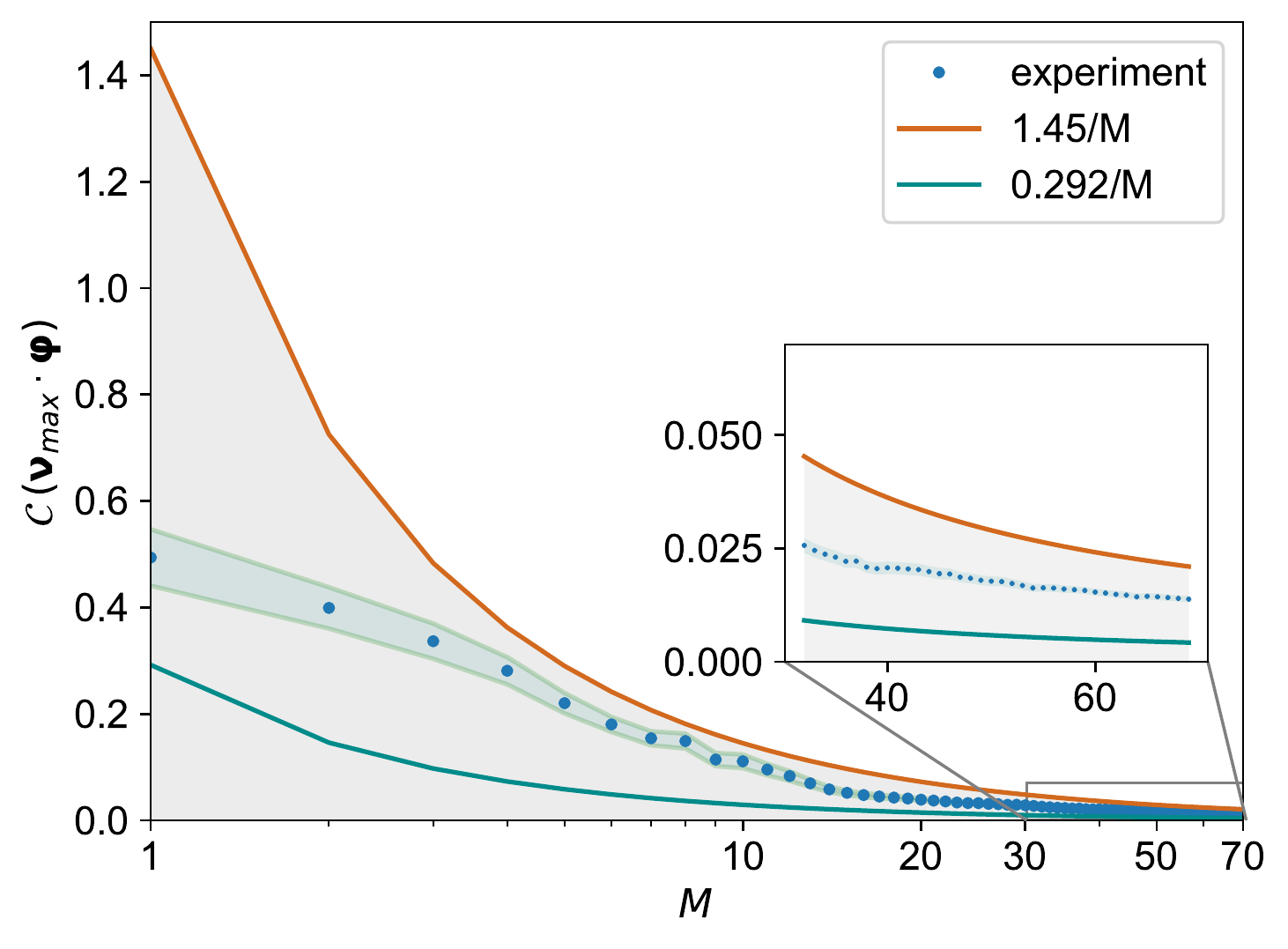}

\caption{\textbf{Experimental adaptive performances.} Estimation performances as a function of the employed resources are reported in terms of $\mathcal{C}(\bm{\nu}_{max}\cdot\bm{\varphi})$. The experimental data with the relative standard deviation (shaded blue region) are averaged over 12 phases estimated 30 independent times. For comparison both the separable (orange line) and parallel (cyan line) bounds are provided. The shaded grey area represents the region showing enhanced sensitivity compared to sequential strategies. The inset shows a zoom of the behaviour for the resources range $M=30$ to $M=70$.}
\label{fig:exp_new} 
\end{figure}

Finally, we also use the adaptive approach to study the estimation of the optimal linear combination of the three parameters discussed in the previous section. The results of the experimental estimates are reported in Fig.\ref{fig:exp_new}. Here, we manage to outperform classical separable strategies, by showing that the average over multiple repetition of the estimation protocol performed on different triplet of phases is found to be below the sequential bound.

%\vspace{1.5cm}

\section*{Discussion}
In this work we have addressed some of the most relevant open issues of multiphase estimation, satisfying simultaneously all the relevant requirements of practical multiparameter quantum metrology in a post-selected configuration.
We demonstrate the saturation of the ultimate precision bound i.e. the QCRB employing multiphoton entangled states.  
We experimentally prove the enhancement achieved using entangled probes over optimal separable estimation strategies when estimating an optimal linear combination of the investigated parameters \cite{malitesta2021distributed}. Furthermore, to grant the optimal sensitivity in the practical limited-resource regime we implement a Bayesian adaptive multiparameter technique,
which requires to operate on a suitably programmable platform applying real-time feedbacks.
We performed our experiment through a versatile setup by means of a state-of-the-art integrated circuit with low insertion losses, low power dissipation and high reliability of the thermal phase shifters, all characteristics that will allow in the future to further scale up the number of spatial modes and the complexity of the devices.
%We provide a demonstration of multiphase estimation simultaneously satisfying, up to employed the post selection, all the relevant requirements of practical quantum metrology. Namely: versatile setup by mean of a state-of-the-art recon- figurable integrated circuit, employment of entangled photon probes, saturation of Quantum Cramer bound, development of Bayesian adaptive multiparameter technique, convergence to the ultimate limits in the limited data regime 

We characterized the integrated circuit using two-photon quantum states and then reconstructing the likelihood function of the device operation. From the collected output statistics, we were able to retrieve the FI matrix of the apparatus, demonstrating the saturation of the QCRB on sensitivity.
Then, we exploited the circuit to perform optimal Bayesian adaptive protocol that allowed us to approach the quantum limits after only $\sim 50$ resources. Notably, the obtained precision  is higher than the one achievable by the best sequential classical strategy  estimating the three phases independently.

The results shown here represent an important step towards the achievement of practical quantum metrology (see Table in the Supplementary Material). 
The demonstrated approach will be the test bed for general quantum multiparameter estimation protocols and promise to host applications like biological sensing \cite{crespi2012measuring}.   
To reach a fully scalable and convenient quantum sensor, two other issues have to be addressed, simultaneously with those closed here. The first one is the scaling of quantum resources: in order to achieve quantum scalings with large number of resources, \cite{cimini2021non} either different kinds of encoding \cite{yokoyama2013ultra} or more efficient sources like quantum dots \cite{moreau2001single} are required. Finally, the unconditional quantum advantage can be claimed if classical limits are overcome even when all the generated resources, including loss and noise mechanisms, are taken into account \cite{slussarenko2017unconditional,zhao2021field}. The most relevant sources of losses lie in the generation and collection of photons. A possible solution to the former is represented by integrated sources \cite{atzeni2018source,paesani2020near} that can be directly interfaced with integrated interferometers. For the detection efficiency, a possible solution is the use of superconductive single photon detectors with near-unity efficiency  \cite{chang2021detecting}.

\section*{Ackwnowledgments}
This work is supported by MIUR (Ministero dell’Istruzione, dell’Università e della Ricerca) via project PRIN 2017 “Taming complexity via QUantum Strategies a Hybrid Integrated Photonic approach” (QUSHIP) Id. 2017SRNBRK, by Ministero dell'Istruzione dell'Universit\`a e della Ricerca (Ministry of Education, University and Research) program ``Dipartimento di Eccellenza'' (CUP:B81I18001170001), by the European Union's Horizon 2020 research and innovation programme under the PHOQUSING project GA no. 899544, and by ERC project CAPABLE ("Composite integrAted Photonic plAtform By ultrafast LasEr micromachining" - Grant Agreement No. 742745). N.S. acknowledges funding from Sapienza Universit\`a via Bando Ricerca 2018: Progetti di Ricerca Piccoli, project "Multiphase estimation in multiarm interferometers". The integrated circuit was partially fabricated at PoliFAB, the micro- and nanofabrication facility of Politecnico di Milano (https://www.polifab.polimi.it/). F.C. and R.O. wish to thank the PoliFAB staff for the valuable technical support.

\bibliographystyle{naturemag}
\bibliography{biblio}

\end{document}

\title{Supplementary Information for experimental multiparameter quantum metrology in adaptive regime}

\author{Mauro Valeri}
\thanks{These authors contributed equally}
\affiliation{Dipartimento di Fisica, Sapienza Universit\`{a} di Roma, Piazzale Aldo Moro 5, I-00185 Roma, Italy}

\author{Valeria Cimini}
\thanks{These authors contributed equally}
\affiliation{Dipartimento di Fisica, Sapienza Universit\`{a} di Roma, Piazzale Aldo Moro 5, I-00185 Roma, Italy}

\author{Simone Piacentini}
\affiliation{Dipartimento di Fisica, Politecnico di Milano, Piazza Leonardo da Vinci, 32, I-20133 Milano, Italy}
\affiliation{Istituto di Fotonica e Nanotecnologie, Consiglio Nazionale delle Ricerche (IFN-CNR), Piazza Leonardo da Vinci, 32, I-20133 Milano, Italy}

\author{Francesco Ceccarelli}
\affiliation{Istituto di Fotonica e Nanotecnologie, Consiglio Nazionale delle Ricerche (IFN-CNR), Piazza Leonardo da Vinci, 32, I-20133 Milano, Italy}

\author{Emanuele Polino}
\affiliation{Dipartimento di Fisica, Sapienza Universit\`{a} di Roma, Piazzale Aldo Moro 5, I-00185 Roma, Italy}

\author{Francesco Hoch}
\affiliation{Dipartimento di Fisica, Sapienza Universit\`{a} di Roma, Piazzale Aldo Moro 5, I-00185 Roma, Italy}

\author{Gabriele Bizzarri}
\affiliation{Dipartimento di Fisica, Sapienza Universit\`{a} di Roma, Piazzale Aldo Moro 5, I-00185 Roma, Italy}

\author{Giacomo Corrielli}
\affiliation{Istituto di Fotonica e Nanotecnologie, Consiglio Nazionale delle Ricerche (IFN-CNR), Piazza Leonardo da Vinci, 32, I-20133 Milano, Italy}

\author{Nicol\`o Spagnolo}
\affiliation{Dipartimento di Fisica, Sapienza Universit\`{a} di Roma, Piazzale Aldo Moro 5, I-00185 Roma, Italy}

\author{Roberto Osellame}
\email{roberto.osellame@cnr.it}
\affiliation{Istituto di Fotonica e Nanotecnologie, Consiglio Nazionale delle Ricerche (IFN-CNR), Piazza Leonardo da Vinci, 32, I-20133 Milano, Italy}

\author{Fabio Sciarrino}
\email{fabio.sciarrino@uniroma1.it}
\affiliation{Dipartimento di Fisica, Sapienza Universit\`{a} di Roma, Piazzale Aldo Moro 5, I-00185 Roma, Italy}

\maketitle

%From this follows that the trasformation of the first quarter is given by: 
%\begin{equation}
%    \begin{bmatrix}
%\sqrt{r_1}\sqrt{r_3}e^{i\phi_2} & i\sqrt{1-r_1}\sqrt{r_3}e^{i\phi_2} & i\sqrt{r_2}\sqrt{1-r_3} & -\sqrt{1-r_2}\sqrt{1-r_3}\\
%i\sqrt{r_1}\sqrt{1-r_3}e^{i\phi_2} & -\sqrt{1-r_1}\sqrt{1-r_3}e^{i\phi_2} & \sqrt{r_2}\sqrt{r_3} & i\sqrt{1-r_2}\sqrt{r_3}\\
%i\sqrt{1-r_1}\sqrt{r_4} & \sqrt{r_1}\sqrt{r_4} & -\sqrt{1-r_2}\sqrt{1-r_4}e^{i\phi_1} & i\sqrt{r_2}\sqrt{1-r_4}e^{i\phi_1}\\
%-\sqrt{1-r_1}\sqrt{1-r_4} & i\sqrt{r_1}\sqrt{1-r_4} & i\sqrt{1-r_2}\sqrt{r_4}e^{i\phi_1} & \sqrt{r_2}\sqrt{r_4}e^{i\phi_1}\\.
%\end{bmatrix}
%\label{quarter}
%\end{equation}

\section{Experimental details}

The complete experimental setup is shown in Fig. 1 of the main text. Single photon-pairs at $785$ nm are generated through a spontaneous parametric down-conversion (SPDC) process by pumping a BBO crystal. The photon-pair is coupled into single-mode fibers directly injected into the chip. The light polarization in such fibers can be controlled in order to correctly prepare the same polarization for both photons. Furthermore, the temporal indistinguishability between the two injected photons is adjusted by means of a delay line (see Supplementary Note 5). %by passing one of the two modes along a delay line according to conventional hong-ou-mandel procedure [ref?]. The traditional dips and peaks of our chip's HOM effect are shown in the Supplementary Information.
Achieving a good value of indistinguishability is critical as it affects the quality of entanglement in the preparation quarter and the achievement of an effective quantum enhancement. Finally, after propagation through the interferometer, the two-photon state is detected employing a probabilistic photon number detection scheme composed of fiber beam splitters (FBSs) along each output mode followed by single-photon detectors (APDs). The presence of FBSs is required to detect the bunching events where two photons exit along the same output port of the integrated interferometer. %almost half of the cases where the two photons take the same chip output. 
Overall, we have a set of $10$ different output probabilities: $4$ related to the events with the photon pair on the same output mode and $6$ events with photons on a different one (see Eq.8 in the main).
%In order to retrieve the correct likelihood behavior, we discard half of the events revealed in the cases of photons along different modes.
We note that measurement of these bunching events is crucial to saturate the QCBR bound.

\section{Fabrication of the device}

The integrated interferometer was fabricated by femtosecond laser writing following the process described in \cite{ceccarelli2020low}. For inscribing single mode waveguides in an alumino-borosilicate glass (EAGLE XG by Corning), a water-immersion $20$x, $0.5$ NA objective was used to focus inside the substrate the beam generated by a Light Conversion Pharos laser system, set to emit pulses at $1030$ nm with a duration of $170$ fs, a repetition rate of $1$ MHz and an energy equal to $290$ nJ. An immersed operation was preferred to a dry process since the optical circuit was inscribed at $30$ \textmu m from the lower surface of the sample, thus a reduction of the spherical aberrations was beneficial for improving the guiding properties. The sample was translated at a speed of 20 mm/s, moreover each waveguide was inscribed with six overlapped laser scans for enhancing the induced refractive index contrast. To reduce both the bending losses and the birefringence of the circuit, a thermal annealing \cite{arriola2013low,corrielli2018symmetric} was performed after the inscription, consisting in a 1 hour long heating ramp up to 750 \textdegree C followed by a slow cooling of the sample, performed with a rate of 12 \textdegree C/h down to 630 \textdegree C, then with a rate of 24 \textdegree C/h down to 500 \textdegree C, and finally by a cooling down to room temperature without thermal actuation. This process provided single mode waveguides at 785 nm, with negligible bending losses at a 30 mm radius of curvature, which was the value chosen in the design of the S-bends. For reducing the footprint of the circuit, the interaction length of the directional couplers was set equal to zero, and a balanced splitting was achieved by employing a coupling distance of 5.67 \textmu m. Finally, the vertical distance between the two waveguides at the crossings in the two quarters was set to $20$ \textmu m, a value sufficient for avoiding an unwanted evanescent coupling. The separation between adjacent modes of the circuits was set to $127$ \textmu m.\\
The thermal phase shifters have been fabricated by depositing by magnetron sputtering a $3$ nm thick adhesion layer of chromium on the whole glass surface, followed by $100$ nm thick layer of gold, which played the role of actual resistive film. In order to increase the stability of the electrical resistivity, and thus the reproducibility of the phase shifting operation, the sample was subjected to a further thermal annealing after the metals deposition. This was composed by a heating ramp of $10$ \textdegree C/min up to $500$ \textdegree C, the maintenance of this temperature for $60$ min and the cooling of the sample down to room temperature without thermal actuation. Subsequently, the pattern of the electrical circuit was defined by femtosecond laser irradiation, selectively ablating the metal layer with a 10x, $0.25$ NA microscope objective focusing pulses at $1030$ nm, $1$ MHz repetition rate, $170$ fs duration, $200$ nJ pulse energy and a sample translation speed of $2$ mm/s. In this way, microheaters with a width of 5 \textmu m and length of 1.5 mm were defined, which show an average resistance value of $110\, \Omega$.
In order to reduce power consumption and mitigate the thermal cross talk between different resistors, thermal insulating trenches were added between adjacent thermal shifters. For this purpose, a water-assisted laser ablation process was exploited \cite{li2013water}, by using the same fabrication setup used for the waveguides inscription (20x, 0.5 NA water-immersion objective, 1030 nm wavelength) but with different pulse parameters (1 ps pulse duration, 1.5 \textmu J energy, 20 kHz repetition rate) and scanning speed (4 mm/s). The result of this process was the ablation of trenches of width equal to 100 \textmu m, length 1.5 mm and depth 300 \textmu m. This process was implemented after waveguide inscription and before metal deposition. The performances of the thermal shifters fabricated in this way are: average power needed for a $2\pi$ phase shift equal of 22 mW, thermal cross talk between neighboring thermal shifters $<20\%$.
Finally, the chip was mounted on an aluminum heat sink, the electrical pads of the resistors were connected by a conductive epoxy glue to two printed circuit boards, and both the input and output optical modes were pigtailed to two fiber arrays (fiber model: 780HP), providing a total insertion loss of 2.5 dB in average for all channels.

\section{Bayesian parameter estimation}

In Bayesian learning, the knowledge on a set of random variables -- the unknown parameters $\bm{\varphi}$ -- is encoded in their probability distribution $p(\bm{\varphi})$. The initial knowledge, i.e. the prior distribution $p_0(\bm{\varphi})$, is updated by the experimental results according to the Bayes' rule. The expected values of the probability distribution provides the estimate $\hat{\bm{\varphi}}$ of the parameters and it is demonstrated to saturate asymptotically the CRB. Also further moments can be computed from $p(\bm{\varphi})$ without the need of additional resources, such as the covariance matrix (see Supplementary Note 4). Here, we employ the algorithm of \cite{Valeri2020} to achieve the QCRB using as few resources as possible. The algorithm is based on a Sequential Monte Carlo technique which approximates the continuous space $\bm{\varphi}$ using a set of discrete particles $\{\bm{\varphi}_i\}$.
The unknown phases $\bm{\varphi}^{(X)}$ are treated as random variables, approximated by a set of particles $\{\bm{\varphi}_i\}$, and their probability distribution $p(\bm{\varphi}^{(X)})$ by $\{p_i\}$, such that $\int p(\bm{\varphi}^{(X)}) d\bm{\varphi}^{(X)} = \sum_{i} p_i = 1$. After each measurement result $\delta$, the Bayesian learning updates the distribution $p(\bm{\varphi}^{(X)})$ with the posterior $p(\bm{\varphi}^{(X)} \vert \delta)$, obtained by the Bayes' rule: $p(\bm{\varphi}^{(X)} \vert \delta) \propto p(\delta \vert \bm{\varphi}^{(X)},\bm{\varphi}^{(C)}) p_0(\bm{\varphi}^{(X)})$. The latter depends on the prior distribution $p_0(\bm{\varphi}^{(X)})$ and the likelihood of the system $p(\delta \vert \bm{\varphi}^{(X)},\bm{\varphi}^{(C)})$, where $\bm{\varphi}^{(C)}$ are parameters controlled for the adaptive estimation.
The estimation $\hat{\bm{\varphi}}^{(X)}$ is provided by the mean value $\hat{\bm{\varphi}}^{(X)} = \int \bm{\varphi}^{(X)} p(\bm{\varphi}^{(X)}) d \bm{\varphi}^{(X)} \approx \sum_i \bm{\varphi}_i p_i$.
Moreover, during the estimation protocol the particles approximation is sometimes adjusted by the resampling technique \cite{granade2012online}, which brings the particles to more likely positions when many of them become no more informative \cite{Valeri2020}. The adaptive techniques allows us to set continuously the device in one of its optimal working points for estimation, tuning the control parameters $\bm{\varphi}^{(C)}$ which set the global phases $\bm{\varphi}^{(X)}+\bm{\varphi}^{(C)}$. According to \cite{Valeri2020}, different reasonable heuristics can be tested to reduce the number of adopted resources. They found the most performing ones based on minimizing the expected overall variance. The strategy corresponds to select the $\bm{\varphi}^{(C)}$ which minimize $\sum_{\delta} p(\delta \vert \bm{\varphi}^{(X)},\bm{\varphi}^{(C)}) \mathrm{Tr}[\mathrm{\Sigma}( \hat{\bm{\varphi}}^{(X)}|\delta,\bm{\varphi}^{(C)})]$, where $\mathrm{\Sigma}( \hat{\bm{\varphi}}^{(X)}|\delta,\bm{\varphi}^{(C)})$ is the covariance that results from the occurrence of outcome $\delta$ for a given $\bm{\varphi}^{(C)}$. The strength of this heuristic relies on its ability to minimize the expected overall variance while taking into account the FI structure, thus finding the right trade-off at any given time. This approach, albeit being computationally expensive, ensures to find optimal controls that perform correctly. The measured performances for such algorithm are shown in Fig.5 of the main text. 

\section{Analysis of the sensor capability}

The performance of the estimation procedure is typically studied in terms of Fisher Information (FI), its inverse is indeed related to the ultimate achievable bound i.e. the Cramér-Rao Bound (CRB). Both these quantities depend on the specific probe employed, the investigated process and finally on the measurement settings whose optimal choice, when possible, allows to saturate the Quantum Cramér-Rao Bound (QCRB). The latter, instead, depends only on the probe state and can be derived by inverting the Quantum Fisher Information (QFI) matrix. In our device, the probe state is generated after the first quarter, therefore, it depends on the selected injection ports and on the values of the phases in the first quarter. The dependence on the phase values of the first quarter of the QCRB when the interferometer is probed by different input states is reported in Supplementary Fig.~\ref{fig:CRB}a, while when injecting the photons either in the first two modes ($|1100\rangle$) or in the last two ($|0011\rangle$) the phase dependence vanishes and the QCRB results to be a constant equal to $2.5$. 
In such configuration the QFI matrix and its inverse read:   

\begin{equation}
\label{eq:QFImatrix}
\text{QFI} = \begin{pmatrix*}[r]
     2 & 0 & -1\\
    0 & 2 & -1\\
    -1 & -1 & 2\\
     \end{pmatrix*} \qquad
\text{QFI}^{-1} = \begin{pmatrix*}[r] 
 0.75 & 0.25 & 0.5\\
0.25 & 0.75 & 0.5\\
0.5 & 0.5 & 1\\ \end{pmatrix*}. \\
\end{equation}

%\begin{figure}[ht!]
%\centering
%\includegraphics[width=.5\textwidth]{QCRB_ideal_1.PNG}
%\caption{QCRB for input states: $|1010\rangle$,$|0101\rangle$,$|0110\rangle$ and $|1001\rangle$.}
%\label{fig:QCRB} 
%\end{figure}

As discussed in the main text, considering two-photon inputs we are interested in three benchmark cases. The lowest bound on the overall estimation error can be achieved when employing entangled states which can be generated injecting into the device two indistinguishable photons. The entangled state is generated after their interference in the first quarter.  In this case, the CRB coincides with the QCRB indeed $\min_{\bm{\varphi}}\mathrm{Tr}[\mathcal{F}^{-1}]=2.5$. The quantum enhancement can be demonstrated computing the same quantity when injecting into the same input ports two completely distinguishable photons for which $\min_{\bm{\varphi}}\mathrm{Tr}[\mathcal{F}^{-1}]=3$. This does not correspond to the optimal classical state that is instead obtained when a superposition of two single photons following Eq.(3) reported in the main text is injected into the device. Such configuration allows to achieve a lower bound $\min_{\bm{\varphi}}\mathrm{Tr}[\mathcal{F}^{-1}]=2.8$. Such conditions can be easily verified looking at Supplementary Fig.~\ref{fig:CRB}b where the CRB obtained when employing $2-$photon indistinguishable and completely distinguishable states are reported and compared to the optimal bounds achievable with classical and quantum states.

%\begin{figure}[ht!]
%\centering
%\includegraphics[width=0.5\textwidth]{CRB_ideal.PNG}
%\caption{$\mathrm{Tr}[\mathcal{F}^{-1}]$ is reported as a function of $\varphi_A$, by fixing $\varphi_B=\varphi_D=0$. In this region, it is visible the minimum CRB achievable by injecting indistinguishable (green) or distinguishable (blue) photons in the device. The dashed gray line represents the ultimate bound achievable with quantum states i.e. $2.5$ while the continuous line represents the one for the optimal classical probe i.e. $2.8$.}
%\label{fig:CRB} 
%\end{figure}

\begin{figure}[ht!]
\centering
\includegraphics[width=\textwidth]{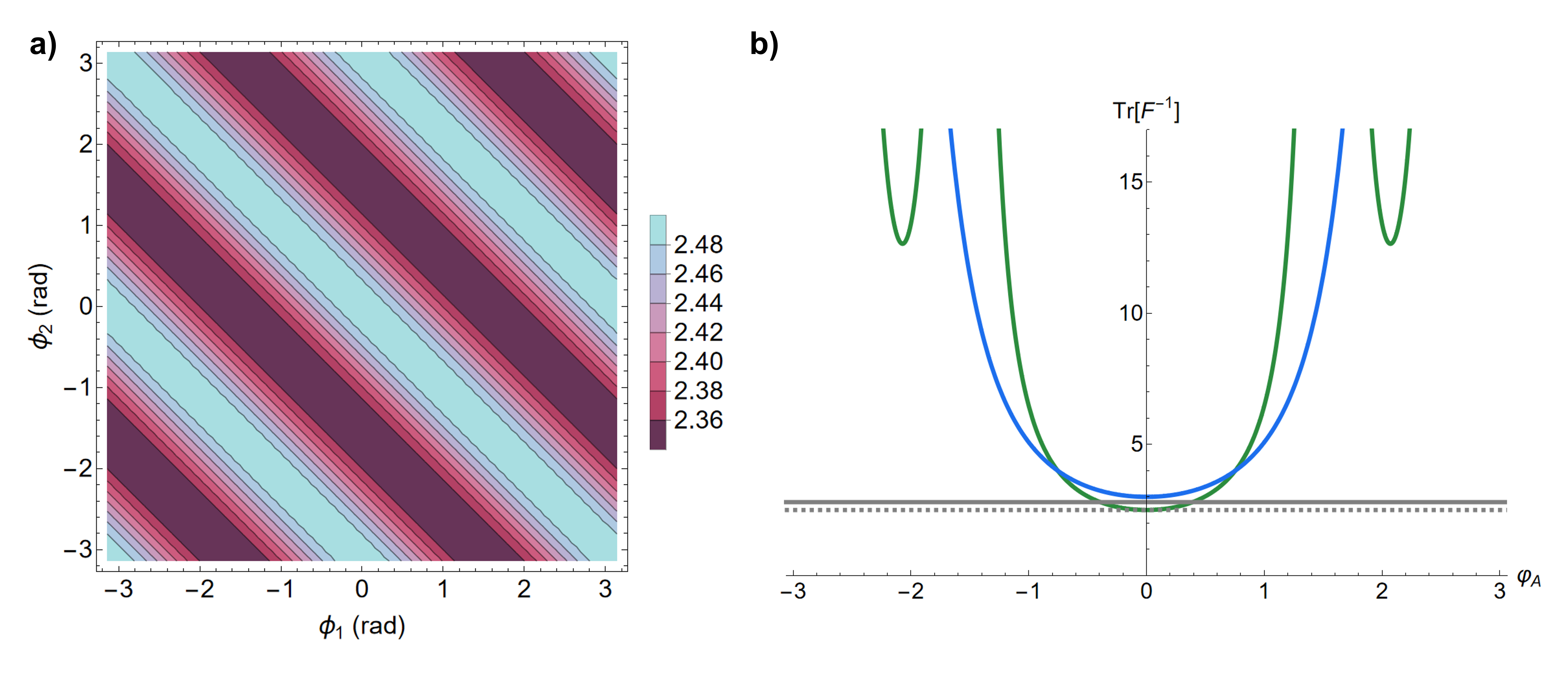}
\caption{\textbf{a)} QCRB for input states: $|1010\rangle$,$|0101\rangle$,$|0110\rangle$ and $|1001\rangle$. \textbf{b)} $\mathrm{Tr}[\mathcal{F}^{-1}]$ is reported as a function of $\varphi_A$, by fixing $\varphi_B=\varphi_D=0$. In this region, it is visible the minimum CRB achievable by injecting indistinguishable (green) or distinguishable (blue) photons in the device. The dashed gray line represents the ultimate bound achievable with quantum states, i.e. $2.5$, while the continuous line represents the one for the optimal classical probe, i.e. $2.8$.}
\label{fig:CRB} 
\end{figure}

%The sensibility of the device in measuring triple phases can be studied in terms of Fisher information matrix $F$. In particular, 
In the indistinguishable case, the bound of $2.5$ for the ideal device is achieved for certain triple of phases. In order to saturate the bound, independently of the triple to be estimated we implement, as explained in the main text, an adaptive algorithm. To understand the working space of the adaptive protocol, it is useful to understand how many points in the whole space allow to reach such ultimate bound.
The trace of the inverse FI matrix provides a benchmark to understand the quality of the specific triple of phases and therefore the working point of the specific estimation. For example, when Tr$(\mathcal{F}^{-1})$ is under a certain threshold $t$ the phases $\bm{\varphi}$ can be estimated with an overall error lower than $t$. Therefore, we compute the density $\rho$ of points under a given $t$ discretizing the whole space with $N$ points and approximating $\rho = \frac{1}{(2\pi)^3} \int_{\text{Tr}(\mathcal{F}^{-1})<t} d\bm{\varphi} \approx \frac{N_t}{N}$, where $N_t$ is the number of discrete points having Tr$(\mathcal{F}^{-1})<t$. We show the behavior of Tr$(\mathcal{F}^{-1})$ in the entire space of phases in Supplementary Fig.~\ref{fig:density}a for the ideal interferometer. The figure shows regions where the Tr$(\mathcal{F}^{-1})$ both is minimal or diverges, comparing the density of points achieving the corresponding bound with the ones of the real device. As expected looking at Fig. 2 in the main text the regions of the real device showing a low value of Tr$(\mathcal{F}^{-1})$ are smaller than the ones of the ideal device.

\begin{figure}[ht!]
\centering
\includegraphics[width=\textwidth]{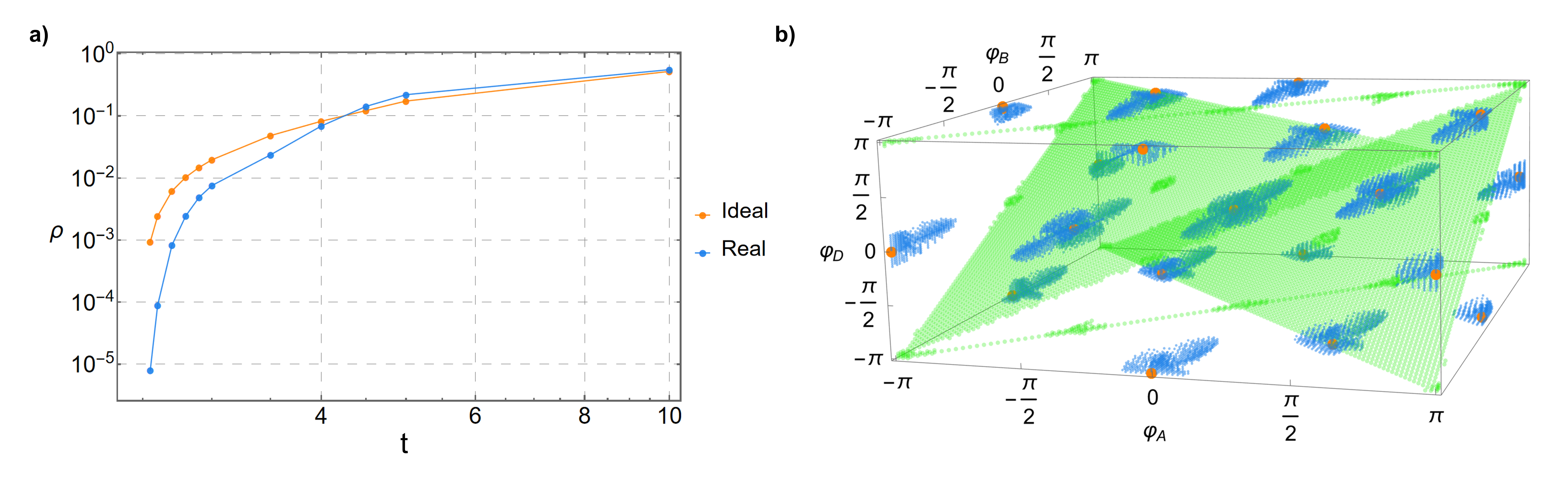}
\caption{\textbf{a)} Number of points normalized to the discretized space related to a Tr$(\mathcal{F}^{-1})<t$, where $t$ is the threshold reported on the $x-$axis. In orange is the CRB related to the ideal device while in blue is the one of the real device. \textbf{b)} The blue points represent the regions corresponding to a value of the CRB $< 2.8$ for the ideal device, already reported in Fig.(2) of the main text. The orange points correspond to the minimum where the CRB$= 2.5$ and therefore the QCRB is saturated. The green points represent instead the divergence regions where Tr$(\mathcal{F}^{-1})\ge 10^4$. }
\label{fig:density} 
\end{figure}

\section{Calibration of the integrated device}

%The optimal functioning of the device is guaranteed when a maximum value of $30$mW of dissipated power on the internal resistors operating at the same time is fulfilled.
The structure of the integrated 4-arm interferometer is reported in Fig. 1 of the main text. It is a highly reconfigurable platform thanks to the presence of several thermo-resistors, controlling both the phases of the external quarters and the ones in the multi-arm interferometer. In order to characterize the device, we start calibrating its response turning on one resistor at a time when injecting single-photon states in each of the device inputs. Only after this initial step we can start to calibrate the cross-talk among different resistors turning on more than one simultaneously. 
In the single-photon regime we collect $4$ output probability related to each of the $4$ different inputs. Once finished the calibration in the single-photon regime we can reconstruct the response function when two-photon states are injected into the device. For such states the possible output configurations for each input state are $10$, as discussed in the main text. To ensure the correct functioning of the device we set as safety operation condition a maximum value of $30$ mW dissipated by each resistor. A $2\pi$ shift can be achieved along a single mode dissipating an electrical power of about $22$ mW on the corresponding resistor. We tune the dissipated power $\omega_k$ changing the electrical current $i_k$ applied to the resistor $k$ we want to work with. In order to retrieve the correct relation we use the following expansions:
\begin{align}
\label{resistances}
\bm{\varphi}_i  = \sum_k (\alpha_{ik}\omega_k + \alpha^{(2)}_{ik}\omega_i\omega_k) + \varphi_{0i}
 \quad,\quad
\omega_k = \frac{R^{(1)}_k}{1-R^{(2)}_k i_k^2}i_k^2.
\end{align}
where the sum is made over the powered resistors, and $\alpha_{ik},\alpha^{(2)}_{ik},R^{(1)}_k,R^{(2)}_k$ are the coefficients to be characterized.
%at the second order of the resistance value:
%\begin{equation}
%    \omega_k \simeq i^2 \frac{R^{(1)}_k}{1-R^{(2)}_k i^2}.
%\label{resistances}
%\end{equation}
Following the same calibration procedure of \cite{Valeri2020}, we identify several static and dynamic parameters. These include the offset phases $\varphi_0$, when no current is applied, relative to the inspected mode, the non-ideal fabrication coefficients of the $8$ directional couplers and the coefficients $\alpha^{(1)}_{ij},\alpha^{(2)}_{ij}$ and $R^{(1)}_k,R^{(2)}_k$ of relations \eqref{resistances}. Furthermore, we consider the visibility $V$ of the two-photon interference and the overall efficiencies on each of the $10$ different output configurations. The final model is retrieved by a fitting procedure with $50$ parameters, giving access to the likelihood function of the device. Examples of the quality of our control of the device are shown in Supplementary Fig.~\ref{fig:prob} which shows a good agreement between the reconstructed model and the experimental results. Using the likelihood function we can therefore compute the actual performances of the device in the 3-phase estimation problem.

\begin{figure*}[h!]
\centering
\includegraphics[width=\textwidth]{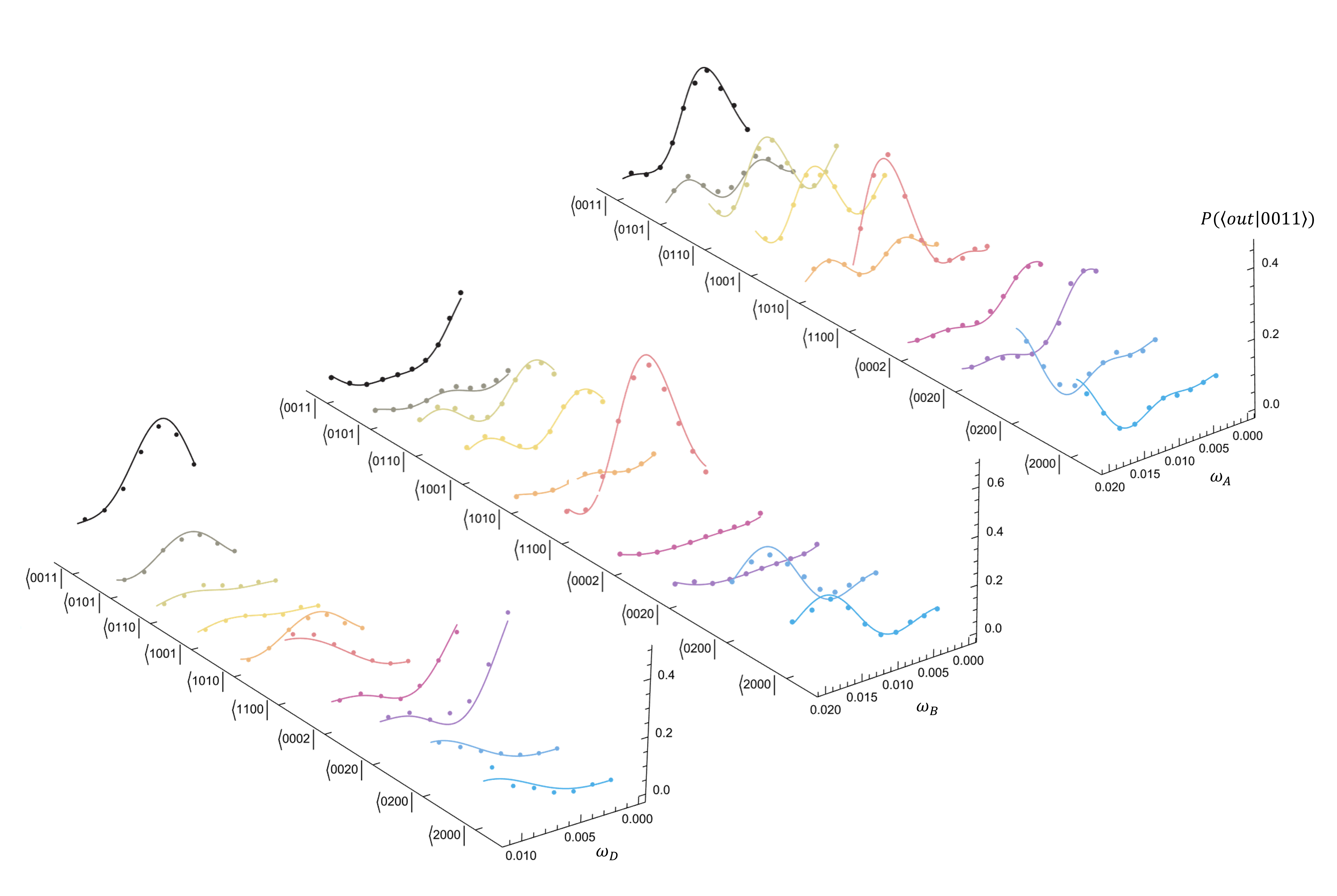}
\caption{Input-output probabilities as a function of the dissipated power on the $3$ resistors employed for the estimation problem. The line represents the probability functions reconstructed via the fitting procedure while the points are the measured probabilities. Error bars are smaller than the points.}
\label{fig:prob} 
\end{figure*}

\section{Preparation of indistinguishable photons}

\begin{figure}[H]
\centering
\includegraphics[width=0.99\textwidth]{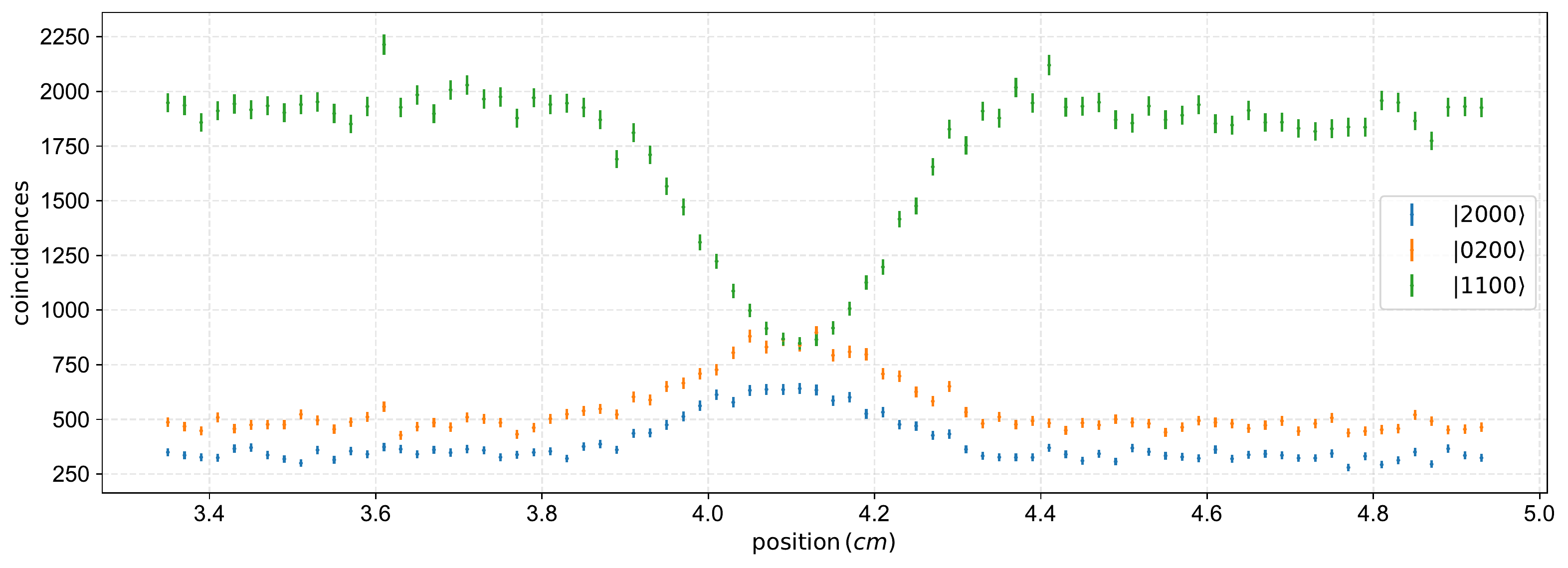}
\caption{Coincidence events between two output modes allowing to measure the degree of distinguishability of the photon pair injected into the device. By tuning the arrival time of the two photons on the chip we transit from distinguishable to indistinguishable photons (center of the coincidence dip/peaks). The arrival time is  set adjusting the position of a mounted delay line along the path of one photon.}
\label{fig:hom_cc}
\end{figure}

Quantum enhancement can be achieved by injecting pair of indistinguishable photons into the device. In order to maximize their temporal indistinguishability we tuned their delay, monitoring the bunching effect \cite{hong1987measurement} at the outputs of the chip. Examples of such effects are shown in Supplementary Fig.~\ref{fig:hom_cc}, where events along the first two output modes are measured. The transition from distinguishable to indistinguishable photons is achieved by means of a delay line that changes the arrival time of the photons at the chip. In the center of the dip, coincidence events of probability $|1100\rangle$ are expected to be half of the distinguishable case, while doubling the bunching events $|2000\rangle$ and $|0200\rangle$.

\section{Comparison with the state of the art}

The implementation of such an effective adaptive protocol in a quantum regime in a multiparameter scenario, is the main result of our work. Thanks to the employment of a tunable integrated photonic chip we are able to continuously set the device in its optimal working point. Therefore, using indistinguishable photon states we were able to implement an estimation protocol of three parameters showing an advantage compared to the estimation performed with classical resources in the limited data regime. Thanks to the use of the highly reconfigurable integrated platform we were indeed able to solve relevant problems, which to the best of our knowledge, still persisted in previous photonic multiphase estimation implementations (as reported in the following Table \ref{tab:multi}).

\begin{table*}[t!]
\centering
\begin{tabular}{|l|l|l|l|l|l|l|l|}
\hline
{\bf Ref.}&{\bf Estimation}&{\bf Probe }&{\bf Quantum } & {\bf  Integrated }&{\bf Saturation}&{\bf Adaptivity} \\
 &{\bf problem}&{\bf }& {\bf probes}& {\bf platform }&{\bf of QCRB}&{\bf } \\
\hline
\cite{polino2019experimental}&Two phases& $N=2$& \textcolor{green}{\cmark} & \textcolor{green}{\cmark}  & \textcolor{red}{\xmark}& \textcolor{red}{\xmark} \\
 && indistinguishable photons&&Three-arm interferometer&&  \\
\hline
 \cite{guo2020distributed} &Average of  &$N\approx 4 \times 2.5$& \textcolor{green}{\cmark}&\textcolor{red}{\xmark}& \textcolor{red}{\xmark}& \textcolor{red}{\xmark}
\\
&four phases  &squeezed coherent state&&4-node bulk network &&
%\\
%\hline
% \cite{zhao2021field} &Difference of  &$N=6$& \textcolor{green}{\cmark}&\textcolor{red}{\xmark}& \textcolor{red}{\xmark}& \textcolor{red}{\xmark}
%\\
%&two phases  &N00N state&&2-node bulk network &&
\\
\hline
 \cite{liu2021distributed} &Average of  &$N=6$& \textcolor{green}{\cmark}&\textcolor{red}{\xmark}& \textcolor{red}{\xmark}& \textcolor{red}{\xmark}
\\
&six phases  &N00N state and multipass&&6-node bulk network &&
\\
\hline
 \cite{roccia2017entangling}&Phase and&$N=2$ & \textcolor{green}{\cmark} & \textcolor{red}{\xmark} & \textcolor{red}{\xmark}& \textcolor{red}{\xmark}
\\
&phase diffusion&separable two-qubit state&&Bulk two-mode &&
\\
&&&&polarization interferometer&&
\\
 \hline
 \cite{Valeri2020}&Two phases & $N=2$& \textcolor{red}{\xmark}& \textcolor{green}{\cmark} & \textcolor{red}{ \xmark}&\textcolor{green}{\cmark}
\\
&&indistinguishable photons&&Three-arm interferometer &&
\\
  \hline
\cite{cimini2019quantum,Roccia18}&Phase and& $N=2$& \textcolor{green}{\cmark}& \textcolor{red}{\xmark} &\textcolor{red}{\xmark}& \textcolor{red}{\xmark} 
\\
&visibility&N00N state &&Bulk two-mode&&
 \\
&&&&polarization interferometer&&
\\
  \hline
\cite{cimini2019adaptive} &Phase and&$N=2$& \textcolor{green}{\cmark}&\textcolor{red}{\xmark} &\textcolor{red}{ \xmark}&\textcolor{green}{\cmark}  
\\
&visibility&N00N state&&Bulk two-mode&Averaged&
 \\
&&&&polarization interferometer&&
\\
\hline
 \cite{parniak2018beating}&Centroid and separation&$N=2$& \textcolor{green}{\cmark} &\textcolor{red}{\xmark} &\textcolor{red}{\xmark}& \textcolor{red}{\xmark}\\
 &of two incoherent sources& indistinguishable photons &  &HOM interference in a BS&&\\
\hline
 \cite{zhou2015quantum}&Unitary process in&$N=4$& \textcolor{green}{\cmark}& \textcolor{red}{\xmark}&\textcolor{red}{\xmark}& \textcolor{red}{\xmark} \\
 &polarization space& Holland-Barnett state &&Bulk polarization&&\\
 &&&&interferometer&&\\
 \hline
 \cite{hong2021quantum}&Three phases&$N=2$& \textcolor{green}{\cmark} &\textcolor{red}{\xmark} &\textcolor{green}{\cmark}& \textcolor{red}{\xmark} \\
&&entangled photons & &Bulk four-arm&&\\
&&&&interferometer&&\\
 \hline
 {\bf this}&Three phases&$N=2$& \textcolor{green}{\cmark}& \textcolor{green}{\cmark} &\textcolor{green}{\cmark}& \textcolor{green}{\cmark} \\
{\bf work}&&indistinguishable photons &&Four-arm interferometer&&\\
 \hline
\end{tabular}
\caption{Table of photonic implementations of multiparameter estimations with the respective accomplished  tasks. For the column "adaptivity" where not specified we refer to an adaptive protocol implemented at each single experiment.} 
\label{tab:multi}
\end{table*}

\providecommand{\noopsort}[1]{}\providecommand{\singleletter}[1]{#1}%